\documentclass[onecolumn,final]{elsart}
\usepackage{amssymb}
\usepackage{amsmath}
\usepackage{bm}
\usepackage[dvips]{graphicx}
\begin{document}
\begin{frontmatter}

\title{Numerical Method \\
	for Hydrodynamic Transport\\
	of Inhomogeneous Polymer Melts}

\author[UCSB_Physics,T-11]{David M. Hall}
\author[T-11]{Turab Lookman}
\author[UCSB_ChemEng]{Glenn H. Fredrickson}
\author[UCSB_ChemEng]{Sanjoy Banerjee}
\address[UCSB_Physics]{ Dept. of Physics, University of California, Santa Barbara, CA 80310}
\address[UCSB_ChemEng]{ Dept. of Chem. Eng. University of California, Santa Barbara, CA 80310}
\address[T-11]{ Theoretical Division, Los Alamos National Laboratory, Los Alamos, NM 87545}

\begin{abstract}

We introduce a mesoscale method for simulating hydrodynamic transport 
and self assembly of  inhomogeneous polymer melts in pressure driven and drag induced flows.
This method extends dynamic self consistent field theory (DSCFT) into the hydrodynamic regime where bulk material transport and viscoelastic effects play a significant role. The method combines four distinct components as a single coupled system, including (1) non-equilibrium self consistent field theory describing block copolymer self-assembly, (2) multi-fluid Navier-Stokes type hydrodynamics for tracking material transport, (3) constitutive equations modeling viscoelastic phase separation, and (4) rigid wall fields which represent moving channel boundaries, machine components, and nano-particulate fillers. We also present an efficient, pseudo-spectral implementation for this set of coupled equations which enables practical application of the model in periodic domains. We validate the model by reproducing well known phenomena including equilibrium diblock meso-phases, analytic Stokes flows, and viscoelastic phase separation of glassy/elastic polymer melts. We also demonstrate the stability and accuracy of the numerical implementation by examining its convergence under grid-size refinement.

\end{abstract}
\end{frontmatter}

\small
\setlength{\parskip}{0cm}
\setlength{\parindent}{0.5cm}
\setlength{\eqntopsep}{5pt}

\section{Introduction}

Many important products are composed of inhomogeneous polymeric fluids which are processed in molten form and subjected to industrial techniques such as melt injection, blow molding, spin casting, and fiber drawing. The physical properties of these products are often dictated by the detailed distribution of their constituent components, which in turn is determined by the manner in which they were processed. Improving the final product may be achieved by improving the process, and computational tools can provide a convenient and cost effective alternative to trial-and-error refinement. At the minimum, such a numerical tool must be capable of simulating phase separation, polymer viscoelasticity, boundary surface wetting, the effects of contaminants, and the manner in which hydrodynamic transport influences these phenomena.

\begin{figure}[tb]
	\center
	\includegraphics[width=0.8\linewidth]{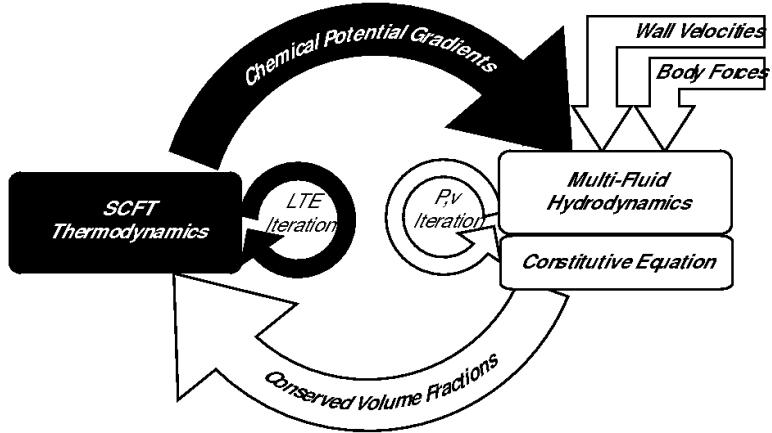}
	\caption{
	Overview of the HSCFT scheme. The thermodynamic model tracks the micro-physical behavior of the polymers, producing imbalanced chemical potential fields which, together with moving walls and external body forces, induce bulk hydrodynamic motion. The multi-fluid model iterates to satisfy the no flow, no slip, constant density, and force balance conditions, and the resultant pressure and velocity fields are used to transport the volume fractions. 
	\label{FIG:OVERVIEW}}
\end{figure}

In this paper, we propose a method for simulating the transport of inhomogeneous polymeric fluids, as well as an efficient pseudo-spectral implementation. The method combines a non-equilibrium version of Self-Consistent Field Theory (SCFT), a Navier-Stokes type hydrodynamic model, and a set viscoelastic constitutive equations, into a single, coupled set of nonlinear differential equations. It also employs a set of continuous, rigid wall fields, which may be moving, to simulate pressure driven flows, drag induced flows, and boundary-wetting conditions. Subsequently, we refer to the method as Hydrodynamic Self Consistent Field Theory (HSCFT) in order to underscore its emphasis on the hydrodynamic transport of polymeric fluids. An schematic diagram illustrating the interaction of each component is presented in fig.\ref{FIG:OVERVIEW}. 

In contrast to ``phase field'' techniques 
\cite{HuoZhangDiblocks,ViscoelasticPhaseSeparation,GersappeShearedBlends} which employ a Ginzburg-Landau free energy, this method is capable of simulating the assembly of multiblock copolymer meso-phases where the polymeric nature of the chains is explicitly taken into account. It also differs from traditional SCFT and  DSCFT methods 
 which are incapable of simulating the effects of hydrodynamic transport
in pressure driven and drag induced flows. Generally speaking, SCFT \cite{CombinatorialScreening,FredricksonBook,MatsenReview} 
describes equilibrium morphologies and meso-phases boundaries, DSCFT \cite{FraaijeMesodyn,ReisterSpinodal,HasegawaDoi,KawakatsuODT,YeungShi,GersappeMicelles} 
models non-equilibrium systems in which hydrodynamic transport may be neglected including phase separating melts and systems subjected to simple shear fields, and HSCFT is appropriate for the description of non-equilibrium systems in which hydrodynamic effects play an important role. 

In section \ref{GOVERNING_EQUATIONS} we present the governing equations for each component of the model. In section \ref{NONDIMENSIONALIZATION}, we convert these equations to a dimensionless form, and in section \ref{NUMERICAL_IMPLEMENTATION} we present a pseudo-spectral numerical method suitable for practical implementation of the method on a computer. We validate the method in section \ref{VALIDATION} by reproducing the results of established numerical techniques and demonstating qualitative agreement with experiment and verify the stability and convergence of the numerical implementation under grid-size refinement.  We summarize our results in section \ref{CONCLUSIONS}. For completeness, we also present derivations of the SCFT thermodynamic model in appendix \ref{THERMO_DERIVATION}  and the multi-fluid hydrodynamic model in appendix \ref{HYDRO_DERIVATION}.

\section{Governing Equations}\label{GOVERNING_EQUATIONS}

\subsection{Polymer Thermodynamics}

The micro-physical, material specific properties of the system are modeled by a non-equilibrium version of SCFT. Self consistent field theory is a mean field, mesoscale technique ideal for simulating the thermodynamic properties of entangled polymers,which is capable of simulating phase separation of incompatible species, polymer-wall interactions, and the effects of polymer polydispersity. Following the formal procedure discussed in appendix \ref{THERMO_DERIVATION}, we obtain the free energy of a blend of $C$ distinct copolymer chains in the canonical ensemble, which takes the form
\begin{eqnarray}
	F = U_{\phi \phi} + U_{\phi \psi} - kT S
\end{eqnarray}
where the enthalpic components of the free energy are
\begin{eqnarray}
	U_{\phi \phi} &=& 
	\frac{kT}{2 v_0} \int \!d\mathbf{r'} \sum_{i=1}^{M}\sum_{j=1}^{M}
	 \phi_i(\mathbf{r'}) \chi_{ij} \phi_j(\mathbf{r'})
	\\
	U_{\phi \psi} &=& 
	\frac{kT}{v_0} \int \!d\mathbf{r'}\sum_{i=1}^{M}\sum_{j=1}^{W} \phi_i(\mathbf{r'}) \xi_{ij} \psi_j(\mathbf{r'})
\end{eqnarray}
The $U_{\phi \phi}$ term represents the energy associated with all two body monomer-monomer interactions, where  $\phi_i(\mathbf{r})$ is the volume fraction of monomers of type $i$ at point $\mathbf{r}$. The sum runs over all $M$ distinct monomer species. Note that for our purposes we consider two species of the same monomer type to be distinct if they belong to different copolymers, as the two types will exhibit different dynamic behavior. The Flory incompatibility parameter,  $\chi_{ij}$, sets the strength of the effective repulsion between them, which is positive for immiscible polymers. Similarly, the $U_{\phi \psi}$ term represents the net energy of all monomer-wall interactions, where $\psi_j(\mathbf{r})$ represents the local volume fraction of solid material of type $j$ out of $W$ possible ``wall" materials.

The free energy also contains the term
\begin{eqnarray}
	-S/k= -\sum_{\alpha=1}^{C} n_\alpha \ln Q_\alpha[\{\omega\}] 
	- \frac{1}{v_0}\int \! d\mathbf{r}' \sum_{i=1}^{M} \omega_i(\mathbf{r}') \phi_i(\mathbf{r}') 
\end{eqnarray}
which describes the net entropy arising from all copolymers in the system. Each term $Q_\alpha$ represents a partition function over all possible configurations of a given copolymer $\alpha$ when subjected to the externally applied fields $\omega_i(\mathbf{r})$ where $i$ runs over all distinct monomers species and the second term ``subtracts off" the effects of the external field. Note that the copolymer index $\alpha$ is a function of monomer index $i$. For example, in a blend containing two triblock copolymer species, $\alpha=1$ for $i\in\{1,2,3\}$ and $\alpha=2$ for $i\in\{4,5,6\}$.

The manner in which the copolymers chains are stretched and arranged in a mean-field sense may be determined by solving two Feynman-Kac style diffusion equations over the polymerization index $s$ for each copolymer $\alpha$,
\begin{eqnarray}
	 \partial_s q_\alpha(\mathbf{r},s) &=& 
	 R^2_{g \alpha} \nabla ^2 q_\alpha(\mathbf{r},s)-
	N_\alpha \Omega_\alpha(\mathbf{r},s) q_\alpha(\mathbf{r},s) 
	 \\
	  \partial_s q^\dagger_\alpha(\mathbf{r},s) &=&
	   R^2_{g \alpha} \nabla ^2 q^\dagger_\alpha(\mathbf{r},s)+
	  N_\alpha \Omega_\alpha(\mathbf{r},s) q^\dagger_\alpha(\mathbf{r},s)
\end{eqnarray}
The function $q(\mathbf{r},s)$ is called the propagator and $q^\dagger(\mathbf{r},s)$ is the co-propagator, where $q(\mathbf{r},s) q^\dagger(\mathbf{r},s)/Q$ represents the probability of finding segment $s$ at the point $\mathbf{r}$ given the initial conditions $q(\mathbf{r},0)=1$ and $q^\dagger(\mathbf{r},1)=1$. The function $\Omega_\alpha(\mathbf{r},s) = \sum_{i=1}^M \omega_i(\mathbf{r})\gamma_i(s)$ is the external field acting on the polymer at $s$, $\gamma_i(s)$ is an occupation function, indicating the linear density of monomer $i$ at index $s$, and $R_{g\alpha}$ is the radius of gyration of polymer $\alpha$ in the absence of external fields. Once the propagator equations have been solved, the partition functions may be computed by averaging over the system volume $V$ for arbitrary $s'$.
\begin{eqnarray}
	Q_\alpha = \frac{1}{V}\int \!d\mathbf{r}' q_\alpha(\mathbf{r}',s')q^\dagger(\mathbf{r}',s')
\end{eqnarray}

Functional differentiation of the free energy with respect to each of the fields $\phi_i(\mathbf{r})$, $\psi_i(\mathbf{r})$, and $\omega_i(\mathbf{r})$ gives the thermodynamic potentials which drive  phase separation in non-equilibrium states.
\begin{eqnarray}
	 \mu_i^\phi(\mathbf{r}) &=& \frac{\delta F}{\delta \phi_i(\mathbf{r})} = 
	 \frac{kT}{v_0}\left(
	 \sum_{j=1}^{M}\chi_{ij}\phi_j(\mathbf{r})
	  + \sum_{l=1}^W \xi_{il}\phi_l(\mathbf{r}) -\omega_i (\mathbf{r}) \right)
	  \\
	  \mu_i^\psi(\mathbf{r}) &=& \frac{\delta F}{\delta \psi_i(\mathbf{r})} = 
	 \frac{kT}{v_0}\left(
	 \sum_{j=1}^{M}\xi_{ij}\phi_j(\mathbf{r})\right)
	 \\
	  \mu_i^\omega(\mathbf{r}) &=& \frac{\delta F}{\delta \omega_i(\mathbf{r})} = 
	 \frac{kT}{v_0}\left( \frac{h_\alpha}{Q_\alpha}\int_0^1 ds\left[
	 q_\alpha(\mathbf{r,s}) q_\alpha^\dagger(\mathbf{r},s) \gamma_i(s)\right] 
	 -\phi_i(\mathbf{r}) \right)
	 \label{MU_OMEGA}
\end{eqnarray}

The chemical potentials $\mu_i^\phi(\mathbf{r})$ and $\mu_i^\psi(\mathbf{r})$ correspond to conserved number density parameters and are typically slow to relax, as transport over large distances may be required to achieve equilibrium. The chemical potential fields $\mu_i^\omega$, on the other hand correspond to the non-conserved, local rearrangement of chains, and as such may be expected to remain close to their equilibrium value, $\mu^\omega(\mathbf{r})=0$. This condition implicitly defines a set of constraints on $\omega_i(\mathbf{r})$, which may be solved numerically, in an iterative manner.

\subsection{Hydrodynamic Transport}
Simulating the flow of inhomogeneous polymeric fluids in industrial processing flows requires a hydrodynamic model that tracks multiple viscoelastic fluids and irregular, possibly moving channel walls.  Both  requirements may be satisfied simultaneously by combining a generalization of the ``two-fluid model" for polymer blends \cite{DynamicCoupling} with flow penalization techniques for irregular boundary surfaces described in \cite{penalizationSchneider}. The interested reader will find the details of this derivation in appendix \ref{HYDRO_DERIVATION}. The resulting multiple-fluid model is composed of a set of modified Navier-Stokes transport equations which relate the forces in the system to the transport of conserved quantities.

In the absence of chemical reactions, the volume fraction fields are conserved as expressed by the continuity equations
\begin{eqnarray}
	\partial_t \phi_i + \nabla \cdot \phi_i \mathbf{v}^\phi_i &=& 0\\
	\partial_t \psi_i + \nabla \cdot \psi_i \mathbf{v}^\psi_i &=& 0
\end{eqnarray}
where $\mathbf{v}^\phi_i(\mathbf{r})$ is the velocity of fluid $\phi_i(\mathbf{r})$ and $\mathbf{v}^\psi_i(\mathbf{r})$ is the velocity of the solid material $\psi_i(\mathbf{r})$. 

It is convenient to divide the velocity of each fluid into collective and relative components,
 $\mathbf{v}^\phi_i = \mathbf{v}_T + \mathbf{w}$. The tube velocity 
 $\mathbf{v}^\psi_T = \Sigma_i^M \alpha_i^\phi \mathbf{v}^\phi_i  + \Sigma_j^W \alpha^\psi_j \mathbf{v}_j^\psi$ represents the collective motion of the network of topological constrains in an entangled polymer melt, as described in Brochard's theory of mutual diffusion \cite{Brochard}, and the stress division parameters $\alpha^\phi$ and $\alpha^\psi$ are obtained by balancing the frictional forces acting on the network as described in \cite{DynamicCoupling}. The quantity $\mathbf{w}_i=\mathbf{v}_i^\phi - \mathbf{v}_T$ represents the velocity of fluid $i$ relative to the network.  

The relative velocity of each component may be obtained from the momentum transport equations in the low Reynold's number limit,
\begin{eqnarray}
	\zeta_i^\phi(\mathbf{v}_i^\phi - \mathbf{v}_T) &=&  
	\alpha^\phi_i \nabla \cdot \bm{\sigma}
	 - \phi_i (\nabla \mu^\phi_i +  \nabla p) + \mathbf{f}^\phi_i 
	 \label{EQN:FLUID_FORCE_BALANCE}
	 \\
	\zeta^\psi_i(\mathbf{v}^\psi_i - \mathbf{v}_T) &=&
	\alpha^\psi_i \nabla \cdot \bm{\sigma} 
	 - \psi_i (\nabla \mu^\psi_i +  \nabla p ) + \mathbf{f}^\psi_i
	 \label{EQN:SOLID_FORCE_BALANCE}
\end{eqnarray}
where $\zeta_i^\phi$ is the friction coefficient associated with fluid $i$ and $\zeta_j^\psi$ is the friction coefficient associated with solid component $j$. Inertial effects are negligible due to the small size and high viscosity of the system and the fluid velocities $\mathbf{v}^\phi$ are maintained in a psuedo-steady state such that frictional forces balance the osmotic forces $\phi_i \nabla \mu^\phi_i$, pressure gradients $\nabla p$, viscoelastic stresses $\nabla \cdot \bm{\sigma}$ and external body forces $\mathbf{f}_i^\phi$. If the velocity field of the solid objects are specified, then $\mathbf{v}^\psi_i(\mathbf{r})$ is known, and Eq.~(\ref{EQN:SOLID_FORCE_BALANCE}) may be solved for $\mathbf{f}_i^\psi(\mathbf{r})$. Integrating this quantity over each rigidly connected region allows us to measure the net force and torque acting on that object.

Summing over the relative transport equations gives the modified Navier-Stokes momentum transport equation in the low Reynolds number limit,
\begin{eqnarray}
	0 = \nabla \pi + \nabla p - \nabla \cdot \bm{\sigma}- \mathbf{f}^\phi - \mathbf{f}^\psi 
	\label{MOMENTUM_BALANCE}
\end{eqnarray}
where we have defined the total osmotic pressure gradient  
$\nabla \pi = \Sigma_i^M \phi_i \nabla\mu^\phi_i + \Sigma_j^W \psi_j \nabla\mu_j^\psi$, the net wall force $\mathbf{f}^\psi(\mathbf{r})=\Sigma_{j=1}^W \mathbf{f}^\psi_j(\mathbf{r})$ and the net body force  $\mathbf{f}^\phi(\mathbf{r})=\Sigma_{i=1}^M \mathbf{f}^\phi_i(\mathbf{r})$. 
This equation may be solved simultaneously with the incompressible continuity equation $\nabla \cdot \mathbf{v}=0$ to obtain the mean velocity field $\mathbf{v}(\mathbf{r})=\Sigma_i \phi_i\mathbf{v}^\phi_i + \Sigma_j \psi_j\mathbf{v}^\psi_j$, and pressure field. Once $\mathbf{v}$ and $\mathbf{w}_i$ are known, the tube velocity may obtained from the relationship.
\begin{eqnarray}
	\mathbf{v}_T = \frac{\mathbf{v} + \sum_{i=1}^M (\alpha_i^\phi - \phi_i)\mathbf{w}_i
	 + \sum_{j=1}^W(\alpha^\psi_j - \psi_j)\mathbf{v}^\psi }
	 {1-\sum_{k=1}^M (\alpha_k^\phi - \phi_k)}
\end{eqnarray}
In the case where the fluids are dynamically matched, the stress division parameters reduce to the volume fractions, and the tube velocity reduces to the mean velocity. 

\subsection{Viscoelastic Constitutive Equations}

Polymers are viscoelastic materials which flow like a liquid over long times and behave like elastic solids when subjected to rapidly changing stresses and strains. Unlike simple Newtonian fluids, the elastic nature of polymers allows them to store energy, resulting in a time dependent stress-strain relationship which is described by the material's constitutive equation. While many constitutive equations are available which approximate these viscoelastic propeties, we have chosen to employ the Oldyrod-B model which allows us to study a purely viscous fluid ($G_i=K_i=0$), a purely viscoelastic melt ($\eta_s=0$), or anything in-between. Following Tanaka's example \cite{ViscoelasticPhaseSeparation}, we employ shear moduli of the form $G_i(\phi) = G_{0i}\phi_i^2$ and bulk moduli of the form $K_i(\phi) = K_{0i}\theta\left( \phi_i - f_i \right)$ where $f_i$ is the average volume fraction of component $i$ in the system and $\theta$ is the step function. These choices enable us to simulate systems composed of materials with a large dynamic contrast such as a glassy elastic polymer blends and facilitates comparison with well established Ginzbug-Landau methods \cite{ViscoelasticPhaseSeparation}.

The total viscoelastic force at point $\mathbf{r}$ is found by summing the elastic forces contributed by each polymeric component and a dissipative viscous force as described by
\begin{eqnarray}
	\nabla \cdot \bm{\sigma}(\mathbf{r}) =
	 \sum_i\nabla \cdot \bm{\sigma}_i(\mathbf{r}) + \eta_s \nabla^2\mathbf{v}_T(\mathbf{r})
\end{eqnarray}
where the shear and bulk stresses stored in component $i$ evolve over time according to the constitutive equation
\begin{eqnarray}
 \overset{\triangledown}{\bm{\sigma}}{}_i &=&
  G_i(\phi) \bm{\kappa}_T 
  + K_i(\phi) (\nabla \cdot \mathbf{v}_T)\bm{\delta}
  - \bm{\sigma}_i/\tau_\alpha 
  \label{ShearStressEv}		 
\end{eqnarray}
Each of the derivatives above is an upper convected derivative,
$ \overset{\triangledown} {\bm{\sigma}} = \partial_t\bm{\sigma}
	+ \mathbf{v}_T \cdot \nabla \bm{\sigma} 
	- \bm{\sigma}\cdot \nabla \mathbf{v}_T 
	- (\nabla \mathbf{v}_T)^\dagger \cdot  \bm{\sigma}$, which tracks the transport of elastic stresses due to fluid convection while eliminating spurious stresses induced by purely rotational motion, and
the quantity
$
	 \bm{\kappa}_T=
	 \nabla \mathbf{v}_T+ (\nabla \mathbf{v}_T)^T
	 -\frac{2}{d}\left( \nabla \cdot \mathbf{v}_T \right)\bm{\delta}
$ is the shear strain-rate tensor.

This constitutive equation is an appropriate for polymeric fluids subjected to moderate strain-rate shearing flows. For simulations with large elastic strains or highly extensional flows, the constitutive equation may be replaced with a more sophisticated phenomenological model or one based upon the SCFT microphysics \cite{FredricksonRheology}.

\section{Nondimensionalization}\label{NONDIMENSIONALIZATION}

We convert the governing equations to a dimensionless form by extracting a characteristic dimensional scale from each quantity.
\begin{eqnarray}
	\begin{array}{ccccc}
		\hat{\mathbf{r}}=\mathbf{r}/L_c \;\;\;&
		\hat{t}=t/t_c \;\;\;&
		\hat{\mathbf{v}}=\mathbf{v}/V_c \;\;\;&
		\hat{F}=F/E_c \;\;\;&
		\hat{\bm{\sigma}}=\bm{\sigma}/\sigma_c
	\end{array}
\end{eqnarray}
The characteristic length-scale $L_c$ is set by the radius of gyration of the smallest unperturbed copolymer chain, $L_c = R_g = b\sqrt{N/2d}$, where $b$ is the statistical segment length, $d$ is the dimension of the system, and $N$ is its polymerization index . The characteristic time-scale is dictated by the smallest reptative disentanglement time $t_c=\min(\tau_\alpha)$, and the characteristic energy is chosen to be the thermal energy, $E_c=kT$. The chartacteristic viscoelastic stress $\sigma_c = \min(G_i)$ is set by the smallest polymer shear modulus, and the characteristic convective velocity is the defined by the ratio $V_c = L_c/t_c$.
With these definitions, the propagator and copropagators equations become
\begin{eqnarray}
	\partial_s q_\alpha &=& +\lambda_\alpha \left[
		 \hat{\nabla}^2 q_\alpha
		- N\Omega_\alpha q_\alpha  \right] 
		\label{eqn:propNondim}\\
	\partial_s q^\dagger_\alpha &=& -\lambda_\alpha \left[
		 \hat{\nabla^2} q^\dagger_\alpha
		- N\Omega_\alpha q^\dagger_\alpha \right]
		\label{eqn:copropNondim}
\end{eqnarray}
where  $\hat{\bm{\nabla}}=(1/L_c)\bm{\nabla}$ and $\lambda_\alpha = N_\alpha/N$, 
and the dimensionless chemical potentials are
\begin{eqnarray}
	\hat{\mu}^{\phi}_i &=&
	\sum_{j=1}^M \chi_{ij} \phi_j
	 +\sum_{k=1}^W \xi_{k i} \psi_k - \omega_i
	  \label{eqn:muPhi_nondimensional}
	\\
	 \hat{\mu}^{\psi}_i &=&
	 \sum_{j=1}^W \chi^\psi_{i j} \phi_j
	  \label{eqn:muPsi_nondimensional}
\end{eqnarray}
The incompressibility condition $\hat{\nabla} \cdot \hat{\mathbf{v}}=0$ 
and continuity equations are unchanged in dimensionless form
\begin{eqnarray}
	\hat{\partial}_t \phi_i&=& 
		- \hat{\nabla} \cdot \phi_i\hat{\mathbf{v}}^\phi_i\\
		\hat{\partial}_t \psi_i&=& 
		- \hat{\nabla} \cdot \psi_i\hat{\mathbf{v}}^\psi_i
\end{eqnarray}
as are the constitutive equations
\begin{eqnarray}
	 \overset{\triangledown}{\hat{\bm{\sigma}}}{}_i &=&
 	 \hat{G}_{0i} \phi^2 \bm{\kappa}_T 
 	 + \hat{K}_{0i}\theta(\phi_i-f_i) (\hat{\nabla} \cdot \hat{\mathbf{v}}_T)\bm{\delta}
 	 - \hat{\bm{\sigma}}_i/\hat{\tau}_\alpha 
\end{eqnarray}
if we define the dimensionless shear moduli as $\hat{G}_{0i}=G_{0i}/\sigma_c$, the dimensionless bulk moduli as $\hat{K}_{0 i}=K_{0 i}/\sigma_c$, and the dimensionless relaxation times as $\hat{\tau}_\alpha=\tau_\alpha/t_c$.
Following convention, we scale each term in the momentum balance equation by the characteristic viscous force $f_c = \eta_c V_c /L_c^2$, where $\eta_c=\sigma_c t_c$, which gives
\begin{eqnarray}
0 = \mbox{Ca}^{-1}\hat{\nabla} \hat{\pi} + \hat{\nabla}\hat{p} 
	 	- \hat{\nabla} \cdot \hat{\bm{\sigma}} 
		 - \hat{\mathbf{f}}^\psi - \hat{\mathbf{f}}^\phi
\end{eqnarray}
where $\mbox{Ca} = \frac{\eta_c v_c}{L_c \mu_c}$ is the Capillary number.
The equations for the relative velocities and wall forces may be expressed 
\begin{eqnarray}	
	 \Gamma \hat{\zeta}^\phi_i \hat{\mathbf{w}}_i  &=&
		 +\alpha^\phi_i \hat{\nabla} \cdot \hat{\bm{\sigma}}
	 	-\phi_i\left(\mbox{Ca}^{-1}\hat{\nabla} \hat{\mu}^\phi_i
	 	+\hat{\nabla} \hat{p} \right)
		- \hat{\mathbf{f}}_i^\phi \\
	\hat{\mathbf{f}}_j^\psi &=&
		- \alpha^\psi_j \hat{\nabla} \cdot \hat{\bm{\sigma}}
	 	+\psi_j \left(\mbox{Ca}^{-1}\hat{\nabla} \hat{\mu}_{j}^\psi
	 	+\hat{\nabla} \hat{p} \right)
	 	+\Gamma \hat{\zeta}^\psi_j(\hat{\mathbf{v}}^\psi_j - \hat{\mathbf{v}}_T)
	 \end{eqnarray}
where $\Gamma = \frac{\zeta_c L_c^2}{\eta_c}$ is a dimensionless friction factor and $\zeta_c = \min({\zeta_{0i}})$ is the smallest monomer friction coefficient. 
For convenience, we will omit the hats from the dimensionless quantities in the subsequent discussion.

\section{Numerical Implementation}\label{NUMERICAL_IMPLEMENTATION}

The simultaneous solution of the thermodynamic, hydrodynamic, and viscoelastic equations requires a fast and efficient implementation to keep the problem numerically tractable. We propose a multi-step strategy with pseudo-spectral spatial discretization and a semi-implicit time discretization which can resolve sharp interfaces with a minimal number of grid points, and lends itself readily to parallelization using freely distributed fftw-mpi fast fourier transform routines. The method is composed of the following major steps:

(1) Solve the $\mu^\omega_i(\mathbf{r})=0$ local thermodynamic equilibrium conditions eq.(\ref{MU_OMEGA}) using an iterative fixed point method and a pseudo-spectral operator splitting scheme to obtain the mean field chemical potentials $\mu^\phi(\mathbf{r})$ and $\mu^\psi(\mathbf{r})$.

(2) Solve the coupled multi-fluid, constitutive equation system using an iterative fixed point method which employs semi-implicit time discretization, pseudo-spectral spatial derivatives, and the Chorin-Temam projection method \cite{ChorinProjection,TemamProjection}, to obtain velocity fields $\mathbf{v}(\mathbf{r})$ and $\mathbf{w}_i(\mathbf{r})$ that simultaneously satisfy  momentum balance eq.(\ref{MOMENTUM_BALANCE}), the divergence free condition $\mathbf{v}=0$, as well as no-flow and no-slip boundary conditions \cite{penalizationAgnot}.

(3) Use the updated velocities to transport the volume fraction fields 
$\phi_i(\mathbf{r})$ and $\psi_j(\mathbf{r})$ in a flux conserving manner.

\subsection{Thermodynamic Iteration}

The chemical potential fields may be obtained by solving local the thermodynamic equilibrium conditions $\mu^\omega(\mathbf{r})=0$ which implicitly specify the conjugate fields $\omega_j(\mathbf{r})$ as a function of the volume fractions $\phi_i(\mathbf{r})$. 
These conditions are highly nonlinear, and are satisfied by an iterative procedure which computes the propagators and adjusts the conjugate fields to reduce the error. 

The propagator diffusion equations are numerically integrated using a well known pseudo-spectral operating splitting technique \cite{PseudospectralMethod} 
\begin{eqnarray}
	q_\alpha(\mathbf{r},s+\Delta s) = 
	e^{- \Omega_\alpha \lambda_\alpha \Delta s/2}
	\mathcal{F}^{-1} \left[e^{- \lambda_\alpha \Delta s k^2}
	\mathcal{F}\left[e^{- \Omega_\alpha\lambda_\alpha \Delta s/2}
	q_\alpha(\mathbf{r},s)\right]\right]
\end{eqnarray}
\begin{eqnarray}
	q^\dagger_\alpha(\mathbf{r},s-\Delta s) = 
	e^{- \Omega_\alpha \lambda_\alpha \Delta s/2}
	\mathcal{F}^{-1} \left[e^{- \lambda_\alpha \Delta s k^2}
	\mathcal{F}\left[e^{- \Omega_\alpha\lambda_\alpha \Delta s/2}
	q^\dagger_\alpha(\mathbf{r},s)\right]\right]
\end{eqnarray}
where $\mathcal{F}$ and $\mathcal{F}^{-1}$ represent forward and inverse Fourier transform operations, and the effective conjugate potential at index $s$ is 
$\Omega_\alpha(\mathbf{r},s) = \sum_i^M \omega_i(\mathbf{r}) \gamma_i(s)$.
The partition function for each copolymer chain
corresponds to the volume average $Q_\alpha = \frac{1}{V}\int q_\alpha(\mathbf{r},1)$
and the auxilliary volume fraction fields 
$\tilde{\phi}_i(\mathbf{r}) = \frac{h_\alpha}{Q_\alpha}\int_0^1 ds q_\alpha(\mathbf{r},s)q_\alpha^\dagger(\mathbf{r},s)\gamma_i(s)$ are computed by 
quadrature over the polymerization index $s$.

The residual errors $\varepsilon_i(\mathbf{r}) = \phi_i(\mathbf{r})-\tilde{\phi}_ i(\mathbf{r})$ are 
reduced by employing a hill-climbing technique with line minimization as described in \cite{CGWithoutPain}. A trial step is taken along the gradient of the energy surface
$\omega_i^\star(\mathbf{r}) = \omega_i^j(\mathbf{r}) + \epsilon \varepsilon_i(\mathbf{r})$, 
 where $\epsilon$ is some small adjustable parameter, and
a second set of errors $\varepsilon^\star_i(\mathbf{r})$ 
is calculated using the updated values. These two points are fit to a parabola to estimate 
the step size $a = \epsilon \left[\frac{\varepsilon_{11}}{\varepsilon_{11}-\varepsilon_{12}}-1\right]$
which minimizes the error in the current search direction, where
 $\varepsilon_{11}=\frac{1}{V}\int d\mathbf{r}\varepsilon_i(\mathbf{r}) \varepsilon_i(\mathbf{r})$ 
 and $\varepsilon_{12}=\frac{1}{V}\int d\mathbf{r}\varepsilon_i(\mathbf{r})\varepsilon^\star_i(\mathbf{r})$.
This value is then used to advance the fields toward a minimal error solution
\begin{eqnarray}
 	\omega_i^{j+1}(\mathbf{r}) = \omega_i^j(\mathbf{r}) + (a-\epsilon) \varepsilon^\star_i(\mathbf{r})
 \end{eqnarray}
Once the residual error has converged to within some acceptable tolerance, 
the chemical potential fields may be calculated using 
eqs. (\ref{eqn:muPhi_nondimensional}) and (\ref{eqn:muPsi_nondimensional}).

\subsection{Hydrodynamic Iteration}

The fields $\bm{\sigma}, \mathbf{v}, \mathbf{w}_i, \mathbf{f}^\psi$ and $p$ are all coupled by the hydrodynamic equations of motion and must be obtained simultaneously using an iterative procedure, which begins by calculating the updated elastic stresses using a semi-implicit scheme in which the nonlinear terms are treated explicitly.
\begin{eqnarray}
	\bm{\sigma}^*_i = 
	\bm{\sigma}^n_i
	+ G_{0i} (\phi_i^n)^2 \bm{\kappa}_T^j
	+ K_{0i}\theta(\phi_i^n - f_i)
	\mathcal{F}^{-1}\left[i\mathbf{k} \cdot \tilde{\mathbf{v}}_T^j\right]\bm{\delta}
\end{eqnarray}
The superscript $n$ counts iteration of the outer, dynamic loop and the superscript $j$ counts the iterations of the inner, psuedo-equilbrium loop. The the linear terms are treated implicitly
and the upper convected derivatives are computed psuedo-spectrally in Fourier space
 \begin{eqnarray}
\bm{\sigma}^j_i =\frac{ \bm{\sigma}^\star 
	-\mathcal{F}^{-1} i\left( 
	 \tilde{\mathbf{v}}_T \cdot \mathbf{k} \tilde{\bm{\sigma}}_i
	 - \mathbf{k} \tilde{\mathbf{v}}_T \cdot \tilde{\bm{\sigma}}_i
	  - \tilde{\bm{\sigma}}_i \cdot \tilde{\mathbf{v}}_T \mathbf{k} \right)}
	{1 + \Delta t/\tau_\alpha}	 
\end{eqnarray}
as is the shear stress tensor
$
	\bm{\kappa}_T^j  = \mathcal{F}^{-1}i\left[ 
	\mathbf{k}\tilde{\mathbf{v}}_T + \tilde{\mathbf{v}}_T\mathbf{k}  \right]
	-(2/d)\mathcal{F}^{-1}\left[i\mathbf{k} \cdot \tilde{\mathbf{v}}_T^j\right]\bm{\delta}
$. The parameter $\Delta t$ is the time step chosen for the outer loop. The divergence of the total viscoelastic stress is also computed psueodospectrally,
$	\nabla \cdot \bm{\sigma}^j = \mathcal{F}^{-1} \left[ i \mathbf{k} \cdot  \sum_i^M
	 \tilde{\bm{\sigma}}_i^j -k^2 \eta_s \tilde{\mathbf{v}}_T^j \right]
$.
The relative velocity fields are computed using the estimated viscoelastic forces
and the pressures produced in the previous iteration,
\begin{eqnarray}
	\mathbf{w}_i^{j+1} =\frac
	 {\alpha_i^{\phi,n} \nabla \cdot \bm{\sigma}^j
	  -\phi_i^n \left[\mbox{Ca}^{-1} \nabla \mu_i^{\phi,n}+\nabla p^j \right]
	  - \mathbf{f}^{\psi,n}_i }
	{\Gamma \zeta_i^n}
\end{eqnarray}
If the walls are externally driven, the motion of each solid object is specified parametrically as a function of time, and the rigid body motions are converted to a set of wall velocity fields $\mathbf{v}_k^{\psi,n}(\mathbf{r})$ which are used to calculate the forces imposed on the fluid by the walls,
\begin{eqnarray}
	\mathbf{f}_k^{\psi,j} =\psi_j^n \left[
		 \mbox{Ca}^{-1}\nabla \mu_k^{\psi,n}
	 	+  \nabla p^j\right]
	 	- \alpha^{\psi,n}_k \nabla \cdot \bm{\sigma}^j
	 	+\Gamma \zeta^{\psi,n}_k({\mathbf{v}}^{\psi,n}_k - \mathbf{v}_T^j)
\end{eqnarray}
The momentum balance condition is satisfied by seeking a steady state solution for
\begin{eqnarray}
	\mathbf{v}^{j+1} =\mathbf{v}^j + h\left(
	 	 \nabla \cdot \bm{\sigma}^j
		- \mbox{Ca}^{-1} \nabla \pi^n
		 - \nabla p^{j+1}
		 + \mathbf{f}^{\psi,j}
		+ \mathbf{f}^{\phi,j}\right)
\end{eqnarray}
over the pseudo-time variable $h$. 

\subsection{Chorin-Temam Projection}

The pressure and velocity fields are obtained simultaneously using the Chorin-Temam projection method \cite{ChorinProjection,TemamProjection}. Gathering the unknown quantities on the left hand side of the equation gives
\begin{eqnarray}
	\mathbf{v}^\star= \mathbf{v}^j
		 + h\left(
		 \nabla \cdot \bm{\sigma}^j
		- \mbox{Ca}^{-1} \nabla \pi^n
		 + \mathbf{f}^{\psi,n}
		 + \mathbf{f}^{\phi,j}
		 \right)
\end{eqnarray}
where $\mathbf{v}^\star = \mathbf{v}^{j+1} + h\nabla p^{j+1}$ may be viewed as the Helmholtz decomposition of a single compressible velocity field.
Taking the divergence of $\mathbf{v}^\star$ and employing the incompressibility condition $\nabla \cdot \mathbf{v}^{j+1}=0$ produces a Poisson equation for the pressure field $ \nabla^2 p^{j+1} = \nabla \cdot \mathbf{v}^\star/h$
which may be solved in Fourier space.
\begin{eqnarray}
	 p^{j+1} =-\mathcal{F}^{-1} \left( i\mathbf{k} \cdot \tilde{\mathbf{v}}^\star / hk^2 \right)
\end{eqnarray}
The divergence-free portion of the velocity field may then be obtained from
\begin{eqnarray}
	\mathbf{v}^{j+1} = \mathbf{v}^\star - h\nabla p^{j+1}
\end{eqnarray}

\subsection{Volume Fraction Transport}

The preceding sequence repeats until the residual errors, $\varepsilon_v = \max \parallel \mathbf{f}_v - \nabla p \parallel$ and $\varepsilon_p = \max \left|\nabla \cdot \mathbf{v} \right|$, are within acceptable tolerances. Once the hydrodynamic iteration has converged, the velocities fields for each component 
$\mathbf{v}_i^{\phi,n+1}= \mathbf{v}_T^{n+1} + \mathbf{w}_i^{\phi,n+1}$ 
are used to transport the volume fraction fields in a flux conserving manner.
\begin{eqnarray}
	\phi^{n+1}_i &=& \phi^n_i 
		+ \Delta t \mathcal{F}^{-1}
		\left[ i\mathbf{k} \cdot \mathcal{F}
		\left[ \phi_i^n \mathbf{v}_i^{\phi,n}\right]\right] \\
	\psi^{n+1}_k &=& \psi^{n}_k 
		+ \Delta t \mathcal{F}^{-1}
		\left[ i\mathbf{k} \cdot \mathcal{F}\left[ \psi_k^{n} \mathbf{v}_k^{\psi,n}\right] \right]
\end{eqnarray}

\subsection{Wall Field Regularization}

Unbounded chemical potentials would be required to produce regions
that are completely free of a given component. Therefore, 
in a manner analogous to other flow penalization techniques
\cite{penalizationSchneider}, we employ slightly porous walls that do not occupy 
the entire volume at a given location. Solid walls with no slip and 
no flow boundary conditions are recovered in the limit 
where the porosity is taken to zero as discussed in \cite{penalizationAgnot}.
For the simulations presented in this paper, we employ a porosity of $\Phi = 0.01$.

Additionally, pseudo-spectral Fourier methods can produce 
unwanted Gibbs phenomena if the system contains overly sharp interfaces.
To prevent this, we apply a Gaussian filter of the form $f(r) = \exp(-r^2/2a)$  to the wall fields $\psi_j(\mathbf{r})$ to smooth the transition from solid regions to liquid regions in the channel. The characteristic width of the gaussian filter, $a$,  employed in our simulations is typically on the order of $0.2R_g$, and the filter is applied by convolving the two functions in Fourier space, $\psi_{j,\mbox{smoothed}} = \mathcal{F}^{-1}(\tilde{\psi}_j \tilde{f})$. The smallest value of alpha which may be employed is dictated by the spatial resolution of the simulation, as the wall transitions must be resolved by several grid-points.

\section{Numerical Tests and Validation}\label{VALIDATION}
We validate the technique by presenting numerical experiments designed to test the capabilities of each model component as well as grid refinement studies which demonstrate the accuracy and stability of the numerical implementation.

 \subsection{Thermodynamic Properties}
 
 \begin{figure}[h]
 
		\resizebox{\linewidth}{!}
		{
 			\includegraphics[width=.25\linewidth]{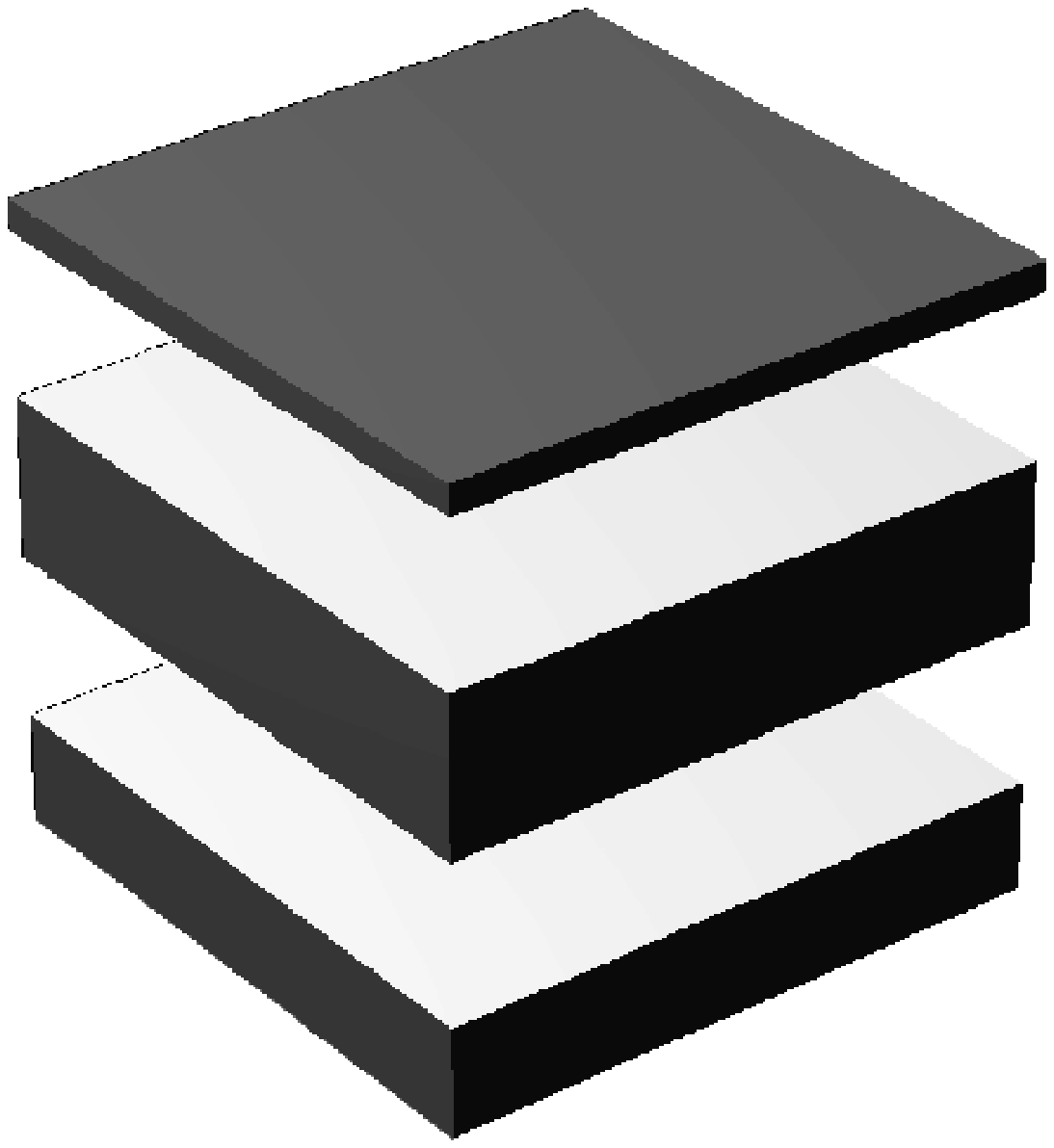}
		 	\includegraphics[width=.25\linewidth]{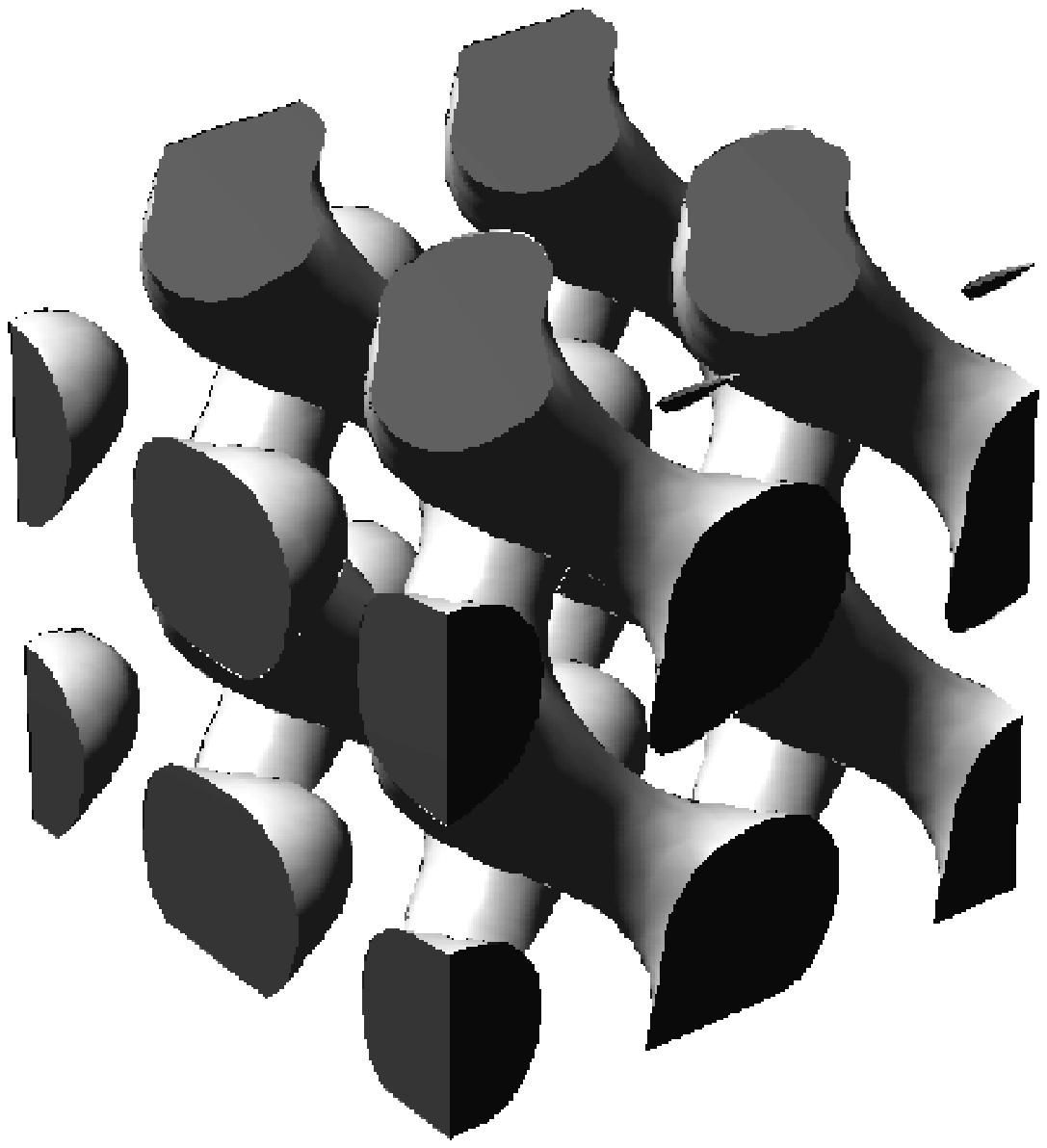}
		 	\includegraphics[width=.20\linewidth]{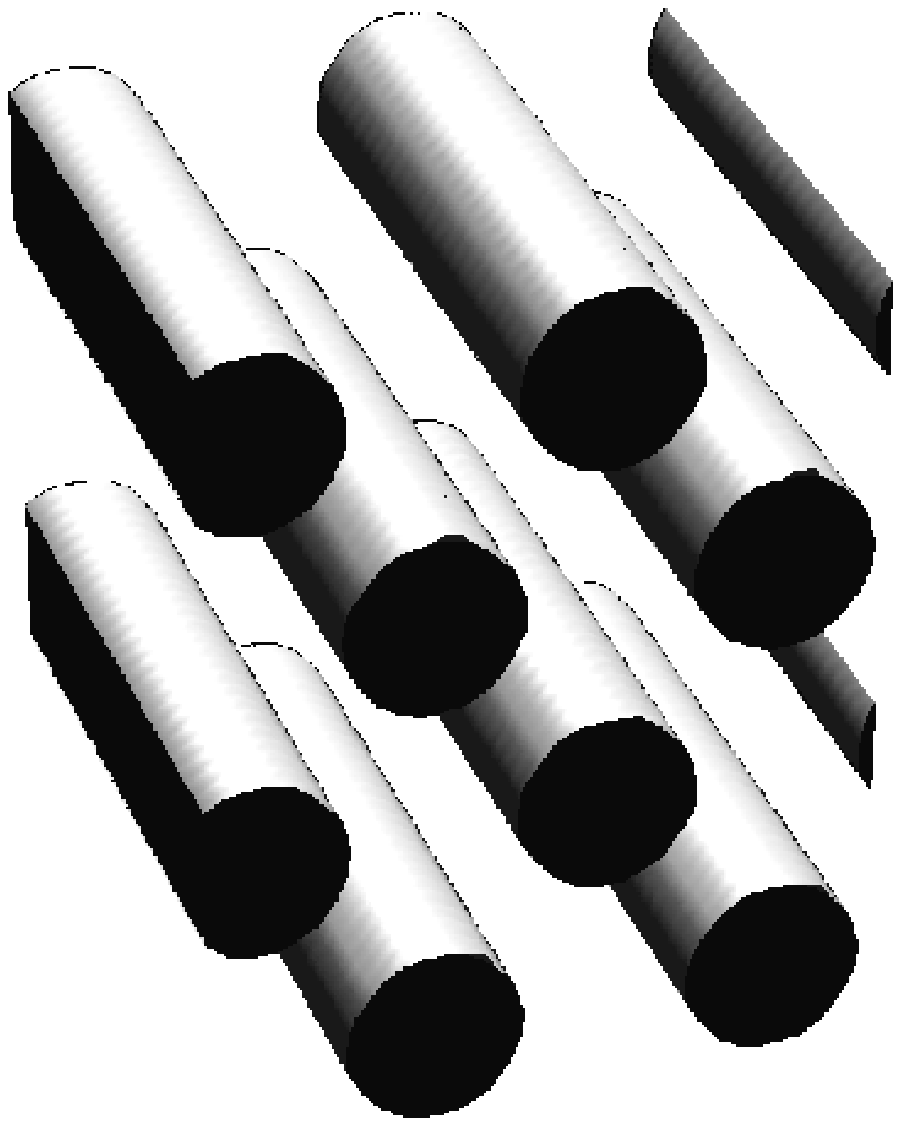}
		 	\includegraphics[width=.25\linewidth]{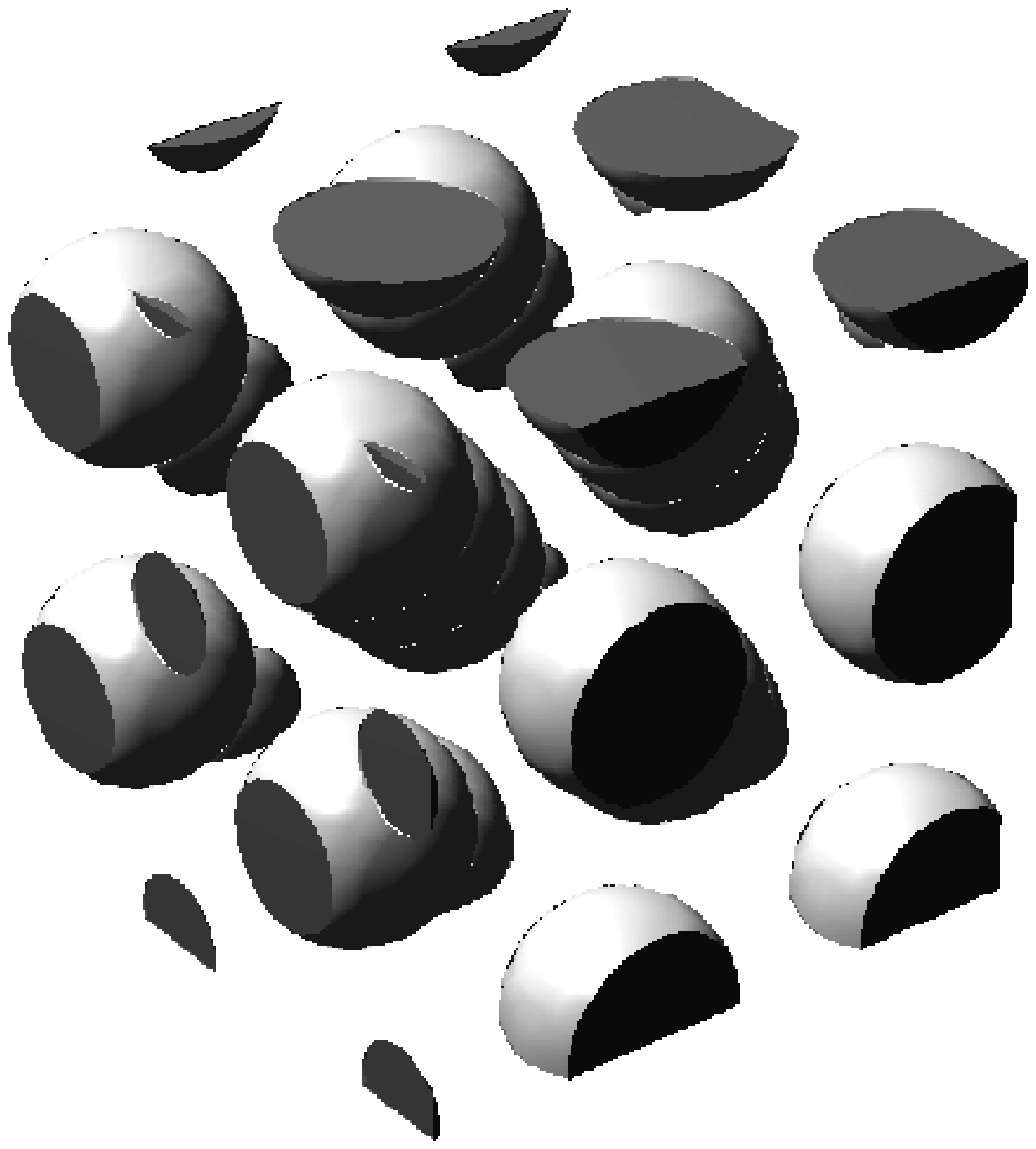}
		}
		 \textbf{a}\hspace{.23\linewidth} \textbf{b}\hspace{.23\linewidth}
		 \textbf{c}\hspace{.18\linewidth} \textbf{d}

		\caption
		{
			\label{DIBLOCK_MESOPHASES}
			Equilibrium diblock copolymer meso-phases are obtained in the long time limit.
			(a) Lamellae, $f_A=0.5$, $\chi N=18$ 
			(b) Gyroid, $f_A=0.37$, $\chi N = 18$ ($Q_{Ia\bar{3}d}$ space group)
			(c) Hexagonally packed cylinders, $f_A=0.3$, $\chi N=21$
			 ($Q_{Im\bar{3}m}$ space group)
			 (d) Close packed spheres, $f_A=0.23$, $\chi N=0.20$
			}
 \end{figure}
 
We validate the thermodynamic components by examining phase separating systems where hydrodynamic and viscoelastic effects play a minimal role.  In particular, we consider the phase separation of diblock copolymer melts in two and three dimensions. This system is of great interest for patterning applications as it exhibits geometrically regular repeating structures with nanometer length scales resulting from the competition between enthalpic and entropic forces. The structure and phase boundaries of these state have been examined in great detail using equilibrium SCFT techniques, and while HSCFT is a non-equilibrium formulation, it should produce the same equilibrium morphologies in the limit of long simulation times.

The equilibrium state exhibited by a particular system is determined by the volume fraction of the components $f_i$ and the degree of incompatibility $\chi N$ between them.  In fig.~\ref{DIBLOCK_MESOPHASES}, we demonstrate that the HSCFT method produces the anticipated equilibrium morphologies in three dimensions including lamellae, gyroids, hexagonally packed cylinders, and close packed spheres. Comparison with the phase boundaries calculated in \cite{MatsenBatesUnifying} confirms that the morphologies obtained are consistent with their coordinates in phase space.

We examine the dynamic evolution of this same system to test the numerical convergence of the technique under grid-size refinement. Consider a symmetric block copolymer melt where both monomer species occupy an equal volume fraction, $f_A=f_B=0.5$, and the Flory interaction strength, $N \chi_{AB}=20$, places the system in the intermediate segregation regime. The melt is initialized to a homogeneous state with small, random fluctuation imposed at all frequencies to break the symmetry.
\begin{figure}[tb]
	\center
		\resizebox{\linewidth}{!}{
		\textbf{a}
		\includegraphics[width=.48\linewidth]{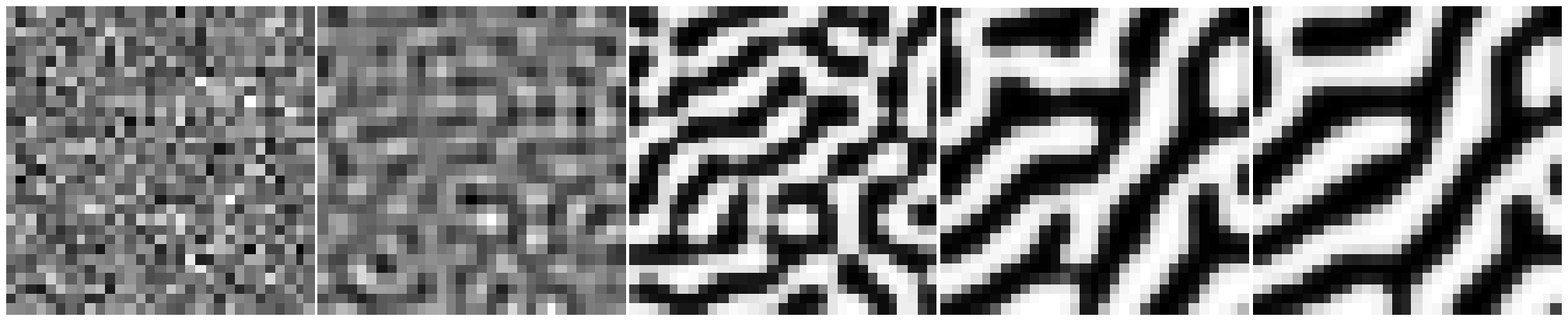}
		\includegraphics[width=.48\linewidth]{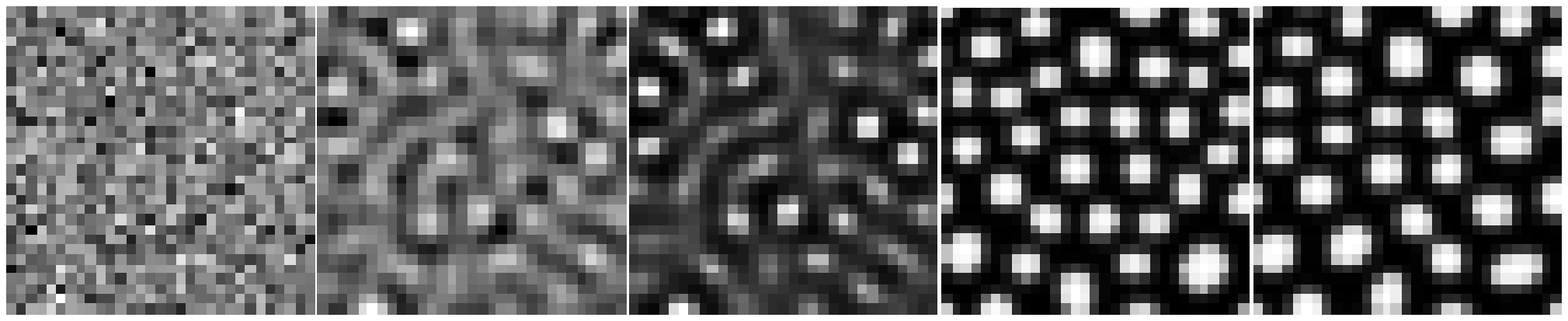}}

		\resizebox{\linewidth}{!}{
		\textbf{b}
		\includegraphics[width=.48\linewidth]{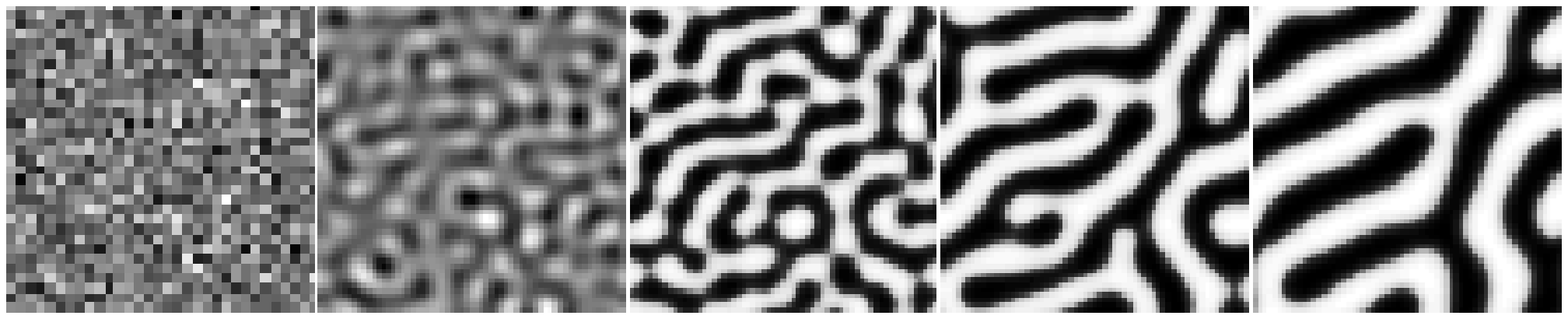}
		\includegraphics[width=.48\linewidth]{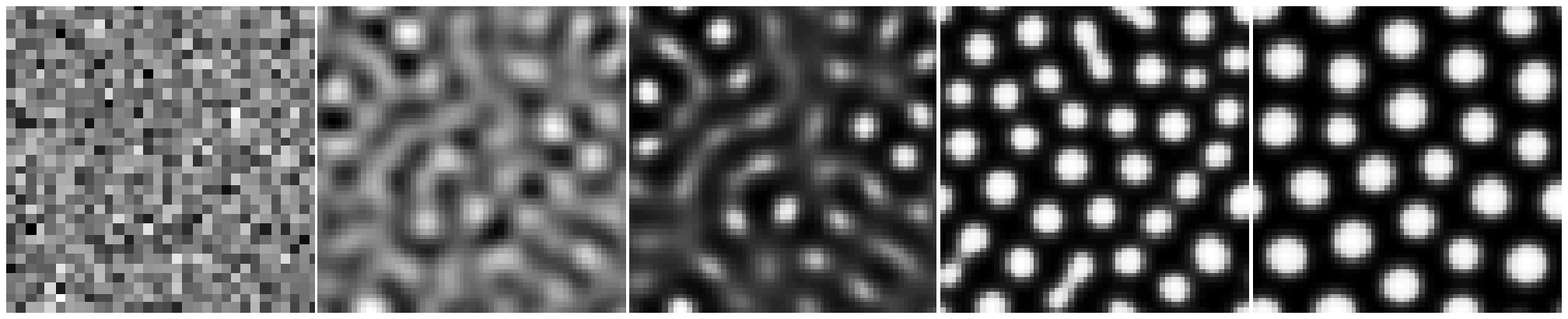}}
		
		\resizebox{\linewidth}{!}{
		\textbf{c}
		\includegraphics[width=.48\linewidth]{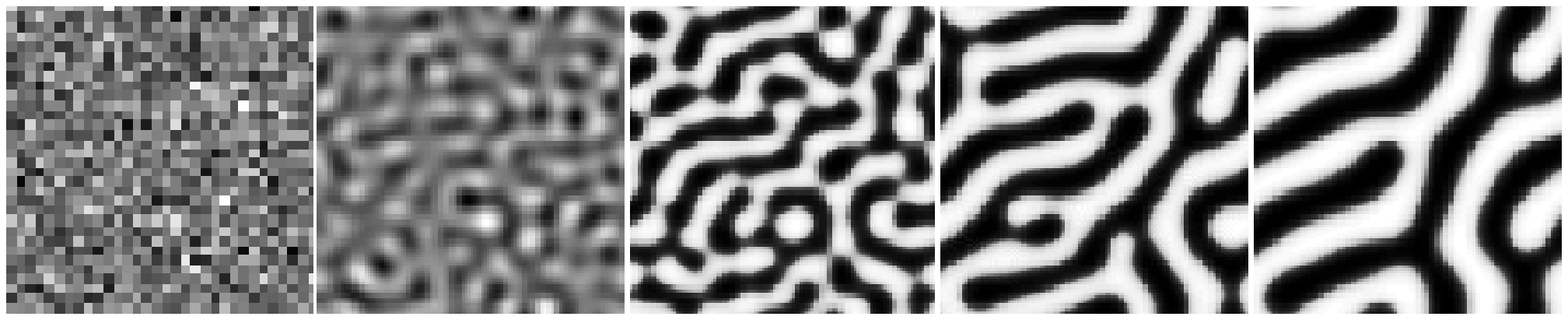}
		\includegraphics[width=.48\linewidth]{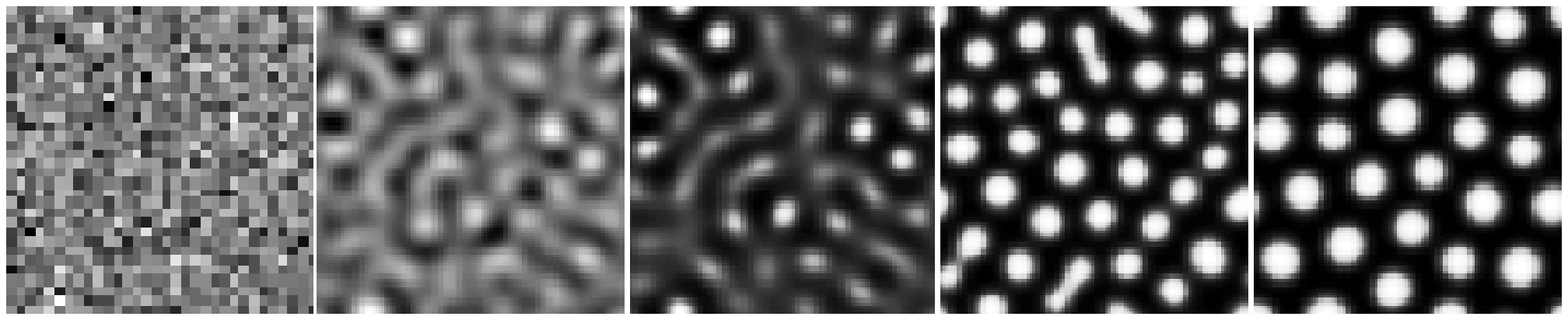}}
		
		\resizebox{\linewidth}{!}{
		\textbf{d}
		\includegraphics[width=.48\linewidth]{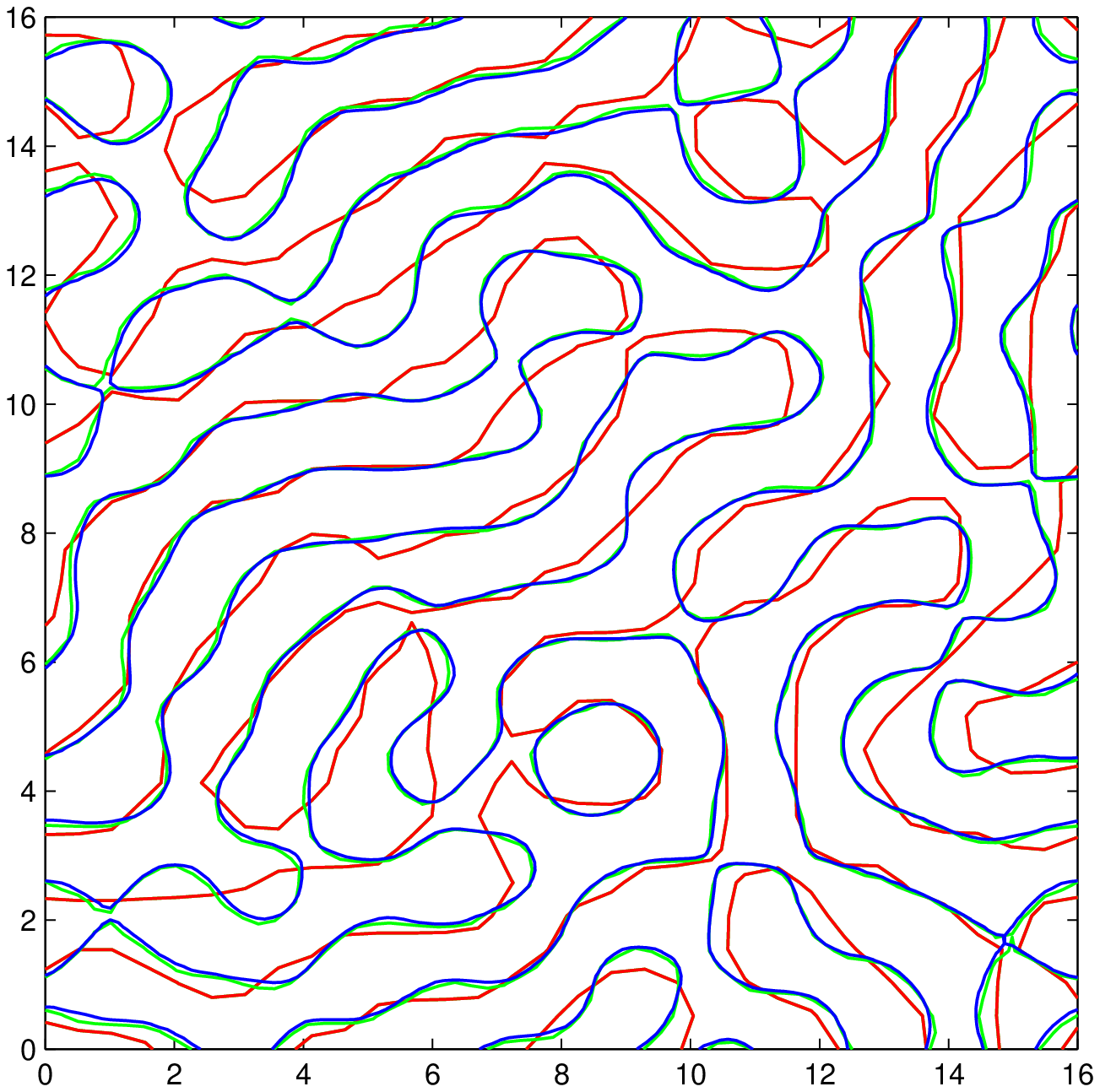}
		\includegraphics[width=.48\linewidth]{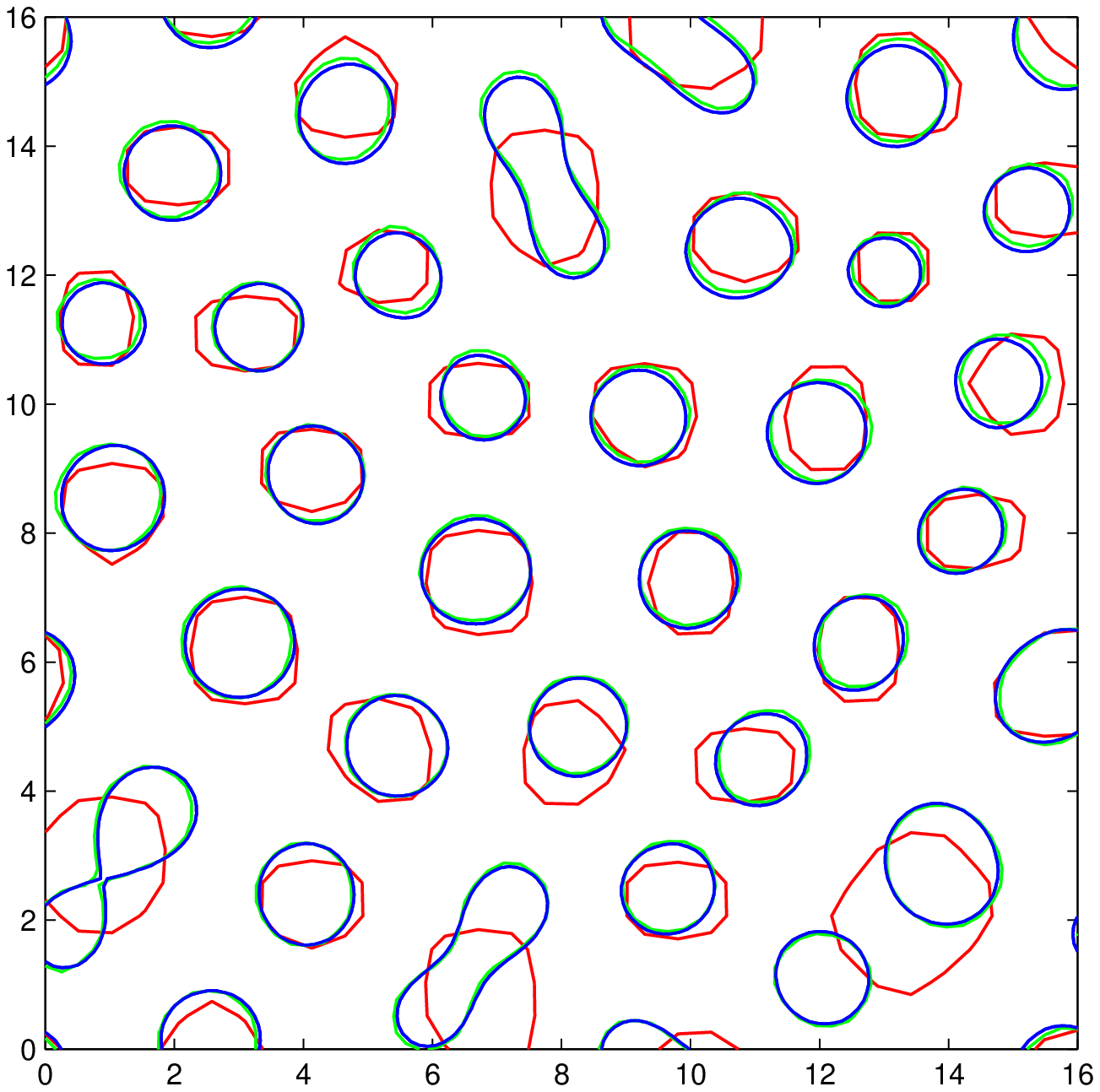}}
	\caption{
	Convergence of diblock copolymer meso-phase separation under grid refinement. 
	[left]: $f_A=0.5,\chi N=20$ at $t= 0,1,3,10,$ and $80$.
	[right]: $f_A=0.3,\chi N=20$ at $t=0,1,4,10$ and $120$.
	Snapshots where taken at three resolutions 
	 (\textbf{a}) 32x32 (red), (\textbf{b}) 64x64 (green), 
	 (\textbf{c}) 128x128 (blue), (\textbf{d}) 
	 Superposition of meso-phase boundaries at all three resolutions at $t=4$.}
	\label{fig:diblock11}
\end{figure}
Fig.~\ref{fig:diblock11} illustrates the time evolution of this system after it has been rapidly quenched below its order disorder transition temperature. Initially, high frequency modes rapidly decay, and density fluctuations grow exponentially, with the greatest growth occurring for structures at a critical wave number. The fluctuations saturate, leaving the melt in a highly defect-filled
meso-phase separated state in which the feature size grows over time, 
until the connectedness of the diblock copolymers prevents further coarsening.
Morphological defects may persist for long times, correspond to kinetically trapped states.

We use the same pattern of random variations about the homogeneous state to initialize simulations of 32x32, 64x64, and 128x128 grid-points in a periodic system of fixed dimension, $16 \times 16 R_g$. When we superimpose the volume fraction contours at all three simulations, we find that the higher resolution simulations converge to a single solution as the mesh is refined. However, the lowest resolution predicts a distinctly different defect pattern
after the fluctuations saturate. We conclude, therefore, that at intermediate segregation strengths,
two gridpoints per $R_g$ is insufficient to fully resolve the mesoscale structures, but four or eight grid points is adequate.

\subsection{Hydrodynamic Properties}

\begin{figure}[h]
	\resizebox{\linewidth}{!}
	{
		\textbf{a} \includegraphics[width=0.25\linewidth]{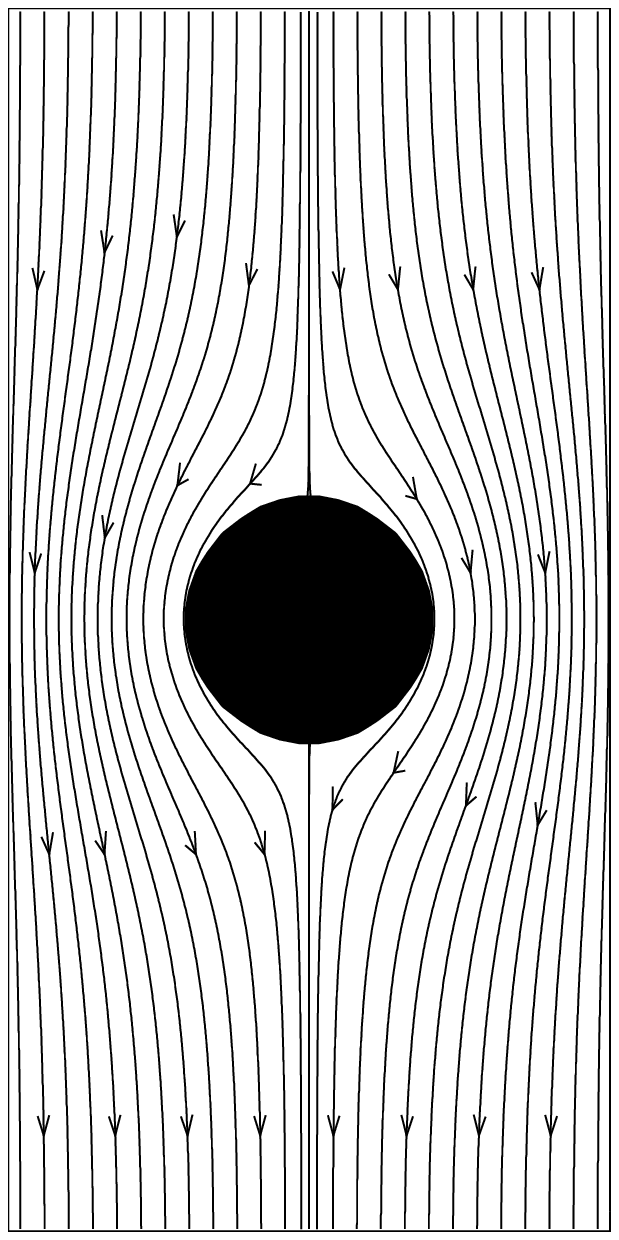}
		\textbf{b} \includegraphics[width=0.25\linewidth]{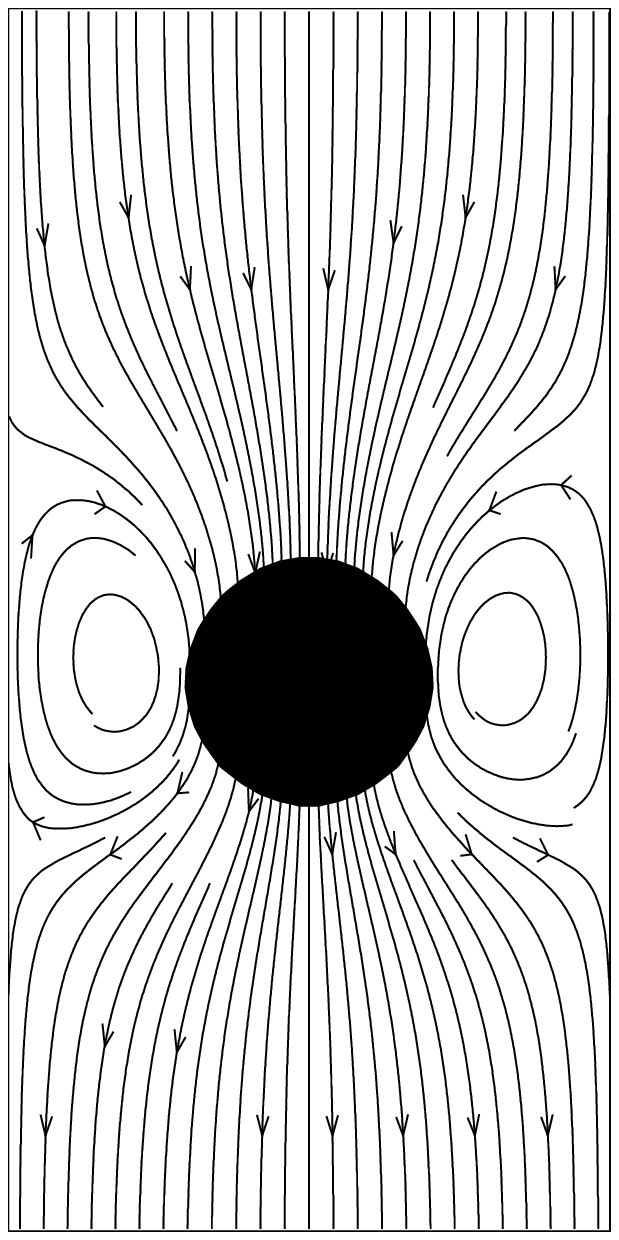}
		\textbf{c} \includegraphics[width=0.50\linewidth]{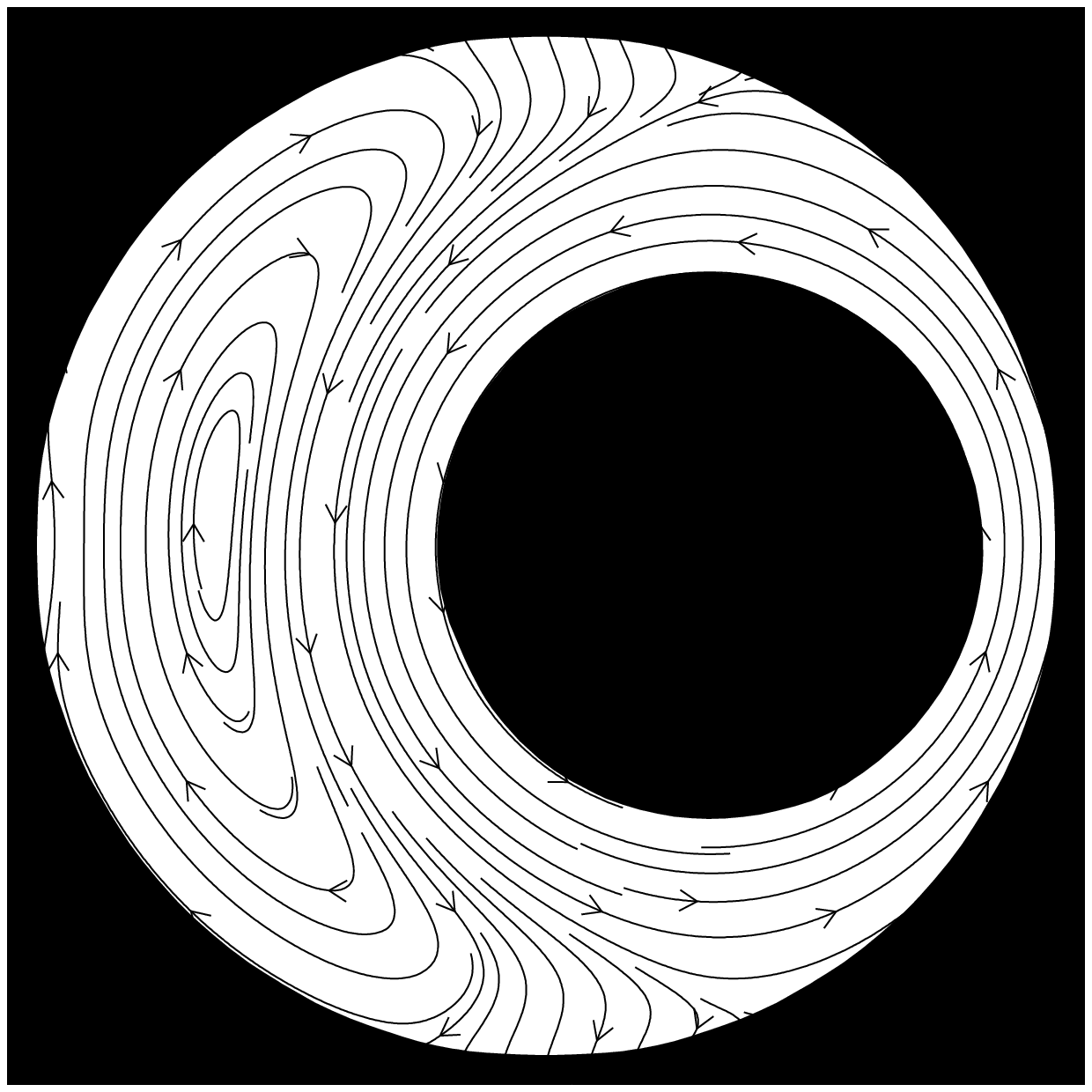}
	}
	\caption{
	Streamlines of wall driven and pressure driven viscous flows at low 
	Reynolds numbers with periodic boundary conditions. 
	(a) Flow past a fixed sphere
	(b) Flow induced by a moving sphere 
	(c) Flow reversal in a journal bearing
	\label{FLOW_VALIDATION}}
\end{figure}

We validate the hydrodynamic component by simulating low Reynolds number, pressure driven and wall driven flows in the absence of thermodynamic or viscoelastic effects. Three classic examples  of Stokes flow found in many texts \cite{Acheson} include flow past a fixed sphere, flow induced by a moving sphere, and the flow generated in a journal bearing due to the motion of two co-rotating eccentric cylinders. The streamlines produced in the simulation of each case is consistent with the known analytical predictions as illustrated in fig.~\ref{FLOW_VALIDATION}. The flow past a fixed sphere (a) exhibits the expected azimuthal symmetry and time reversibility. The streamlines diverge as they flow past the fixed sphere (a) and converge in the presence of the moving sphere in fig.~\ref{FLOW_VALIDATION}(b). The third case (c) corresponds to the counter clockwise rotation of an infinite cylinder in a fixed cylindrical cavity, which produces a counter clockwise rotation in the region closest to the rotating cylinder and a reversed, clockwise flow in the regions furthest from it. This result is also consistent with analytic solutions of the Stokes equations \cite{Acheson} as well as with experimental observations \cite{Chaiken}.

\begin{figure}[htb]
	\center
	\begin{minipage}[b]{0.60\linewidth}
		\textbf{a}\includegraphics[width=\linewidth]{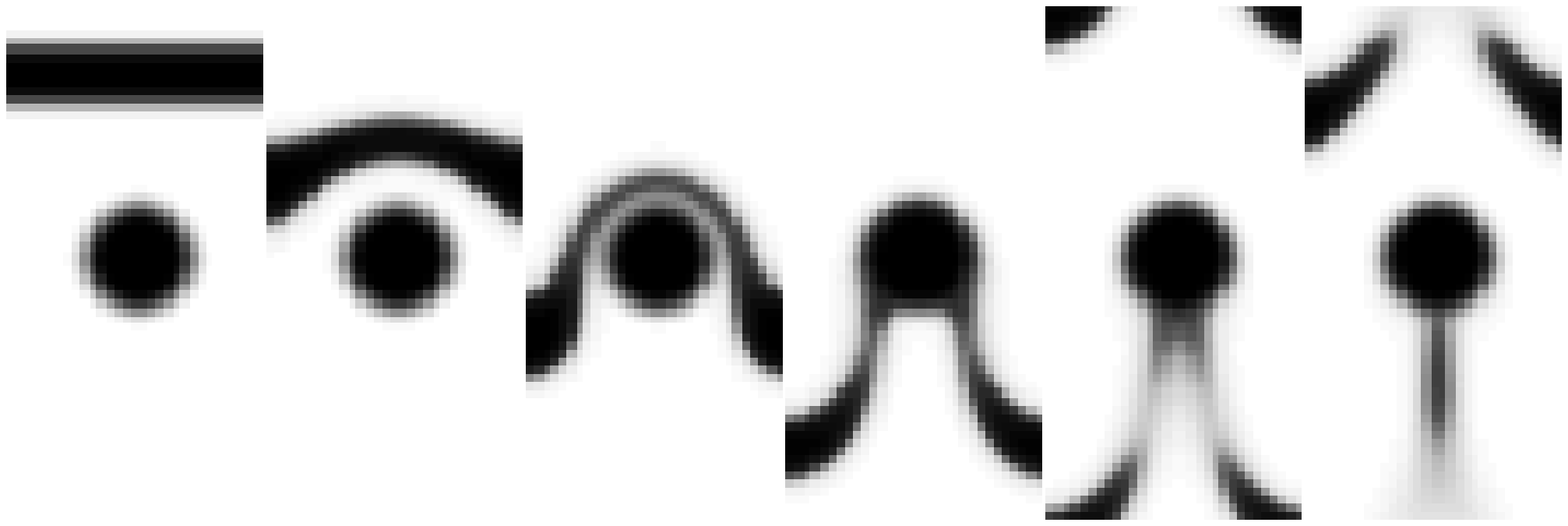}
		\textbf{b}\includegraphics[width=\linewidth]{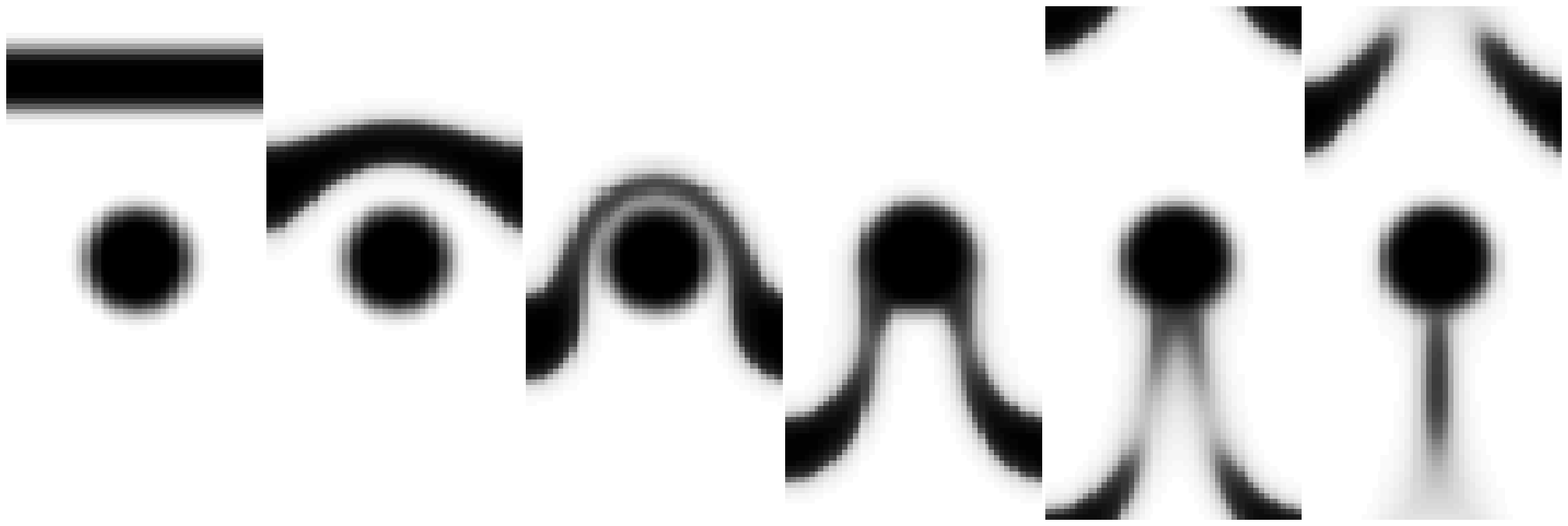}
		\textbf{c}\includegraphics[width=\linewidth]{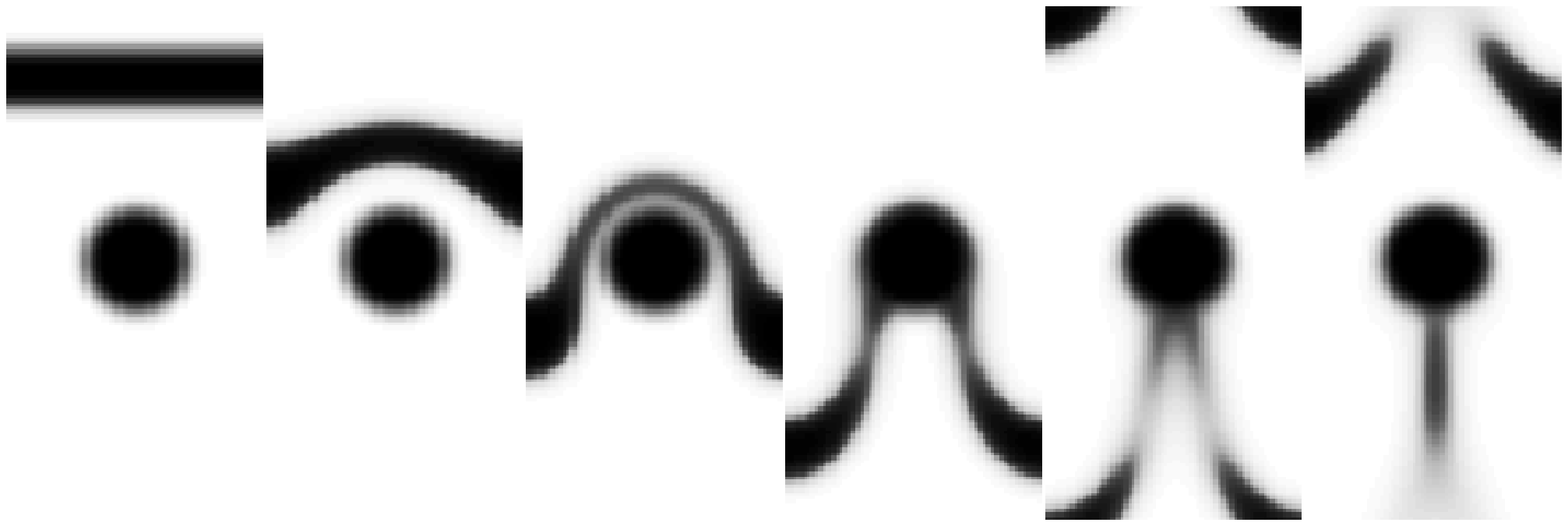}
	\end{minipage}
	\textbf{d}\includegraphics[width=.33\linewidth]{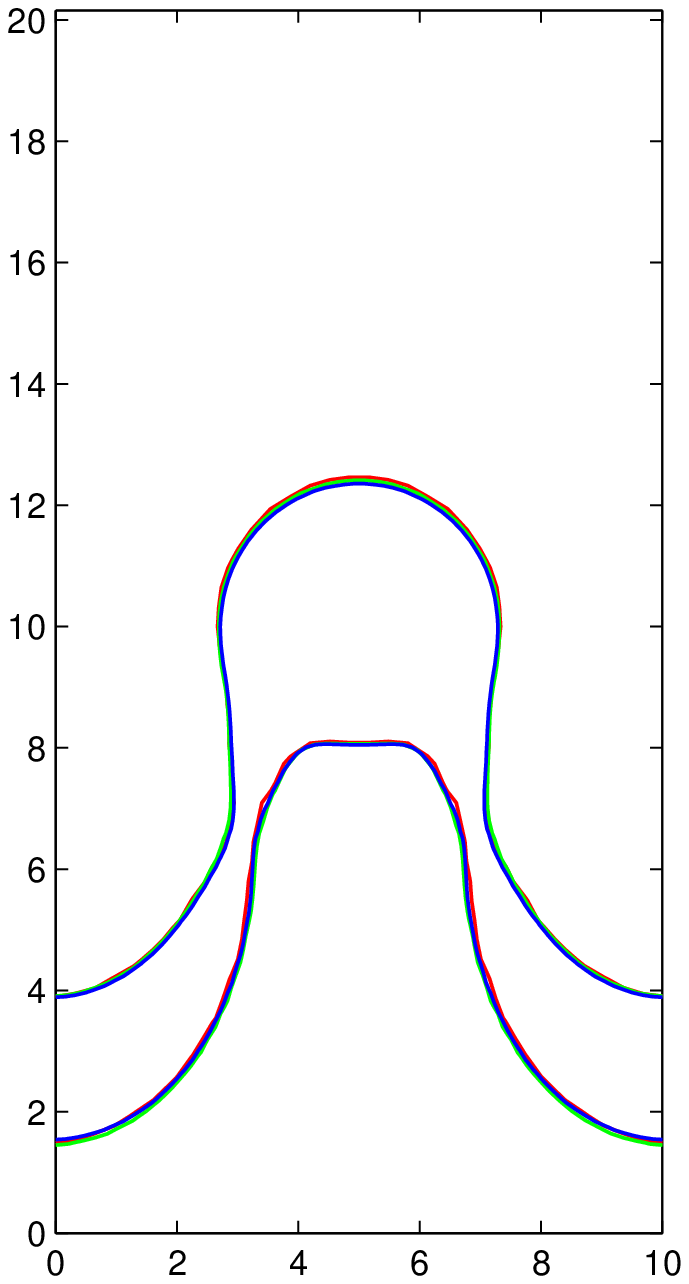}

	\textbf{e}
	\includegraphics[width=0.48\linewidth]{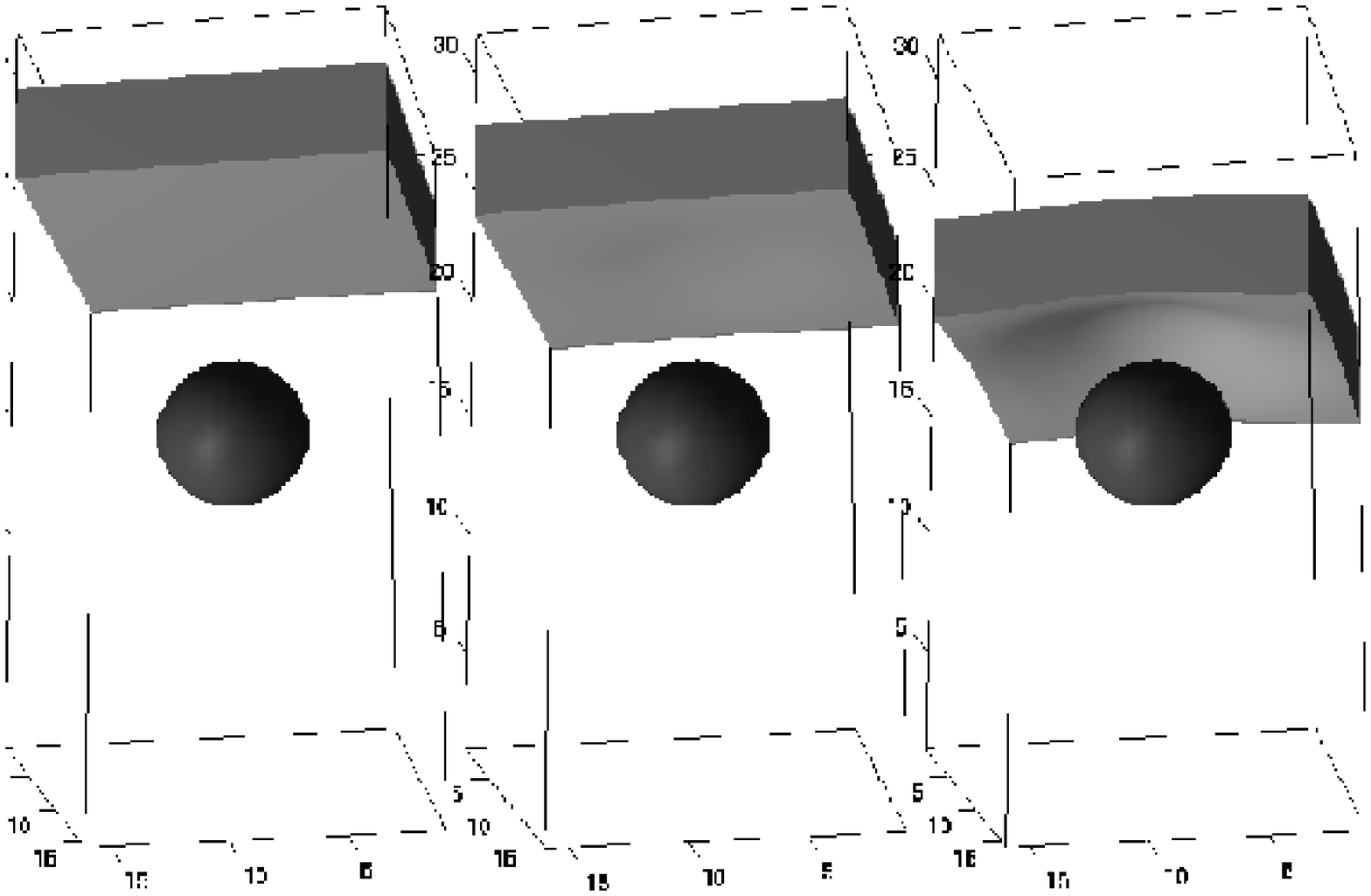}
	\includegraphics[width=0.48\linewidth]{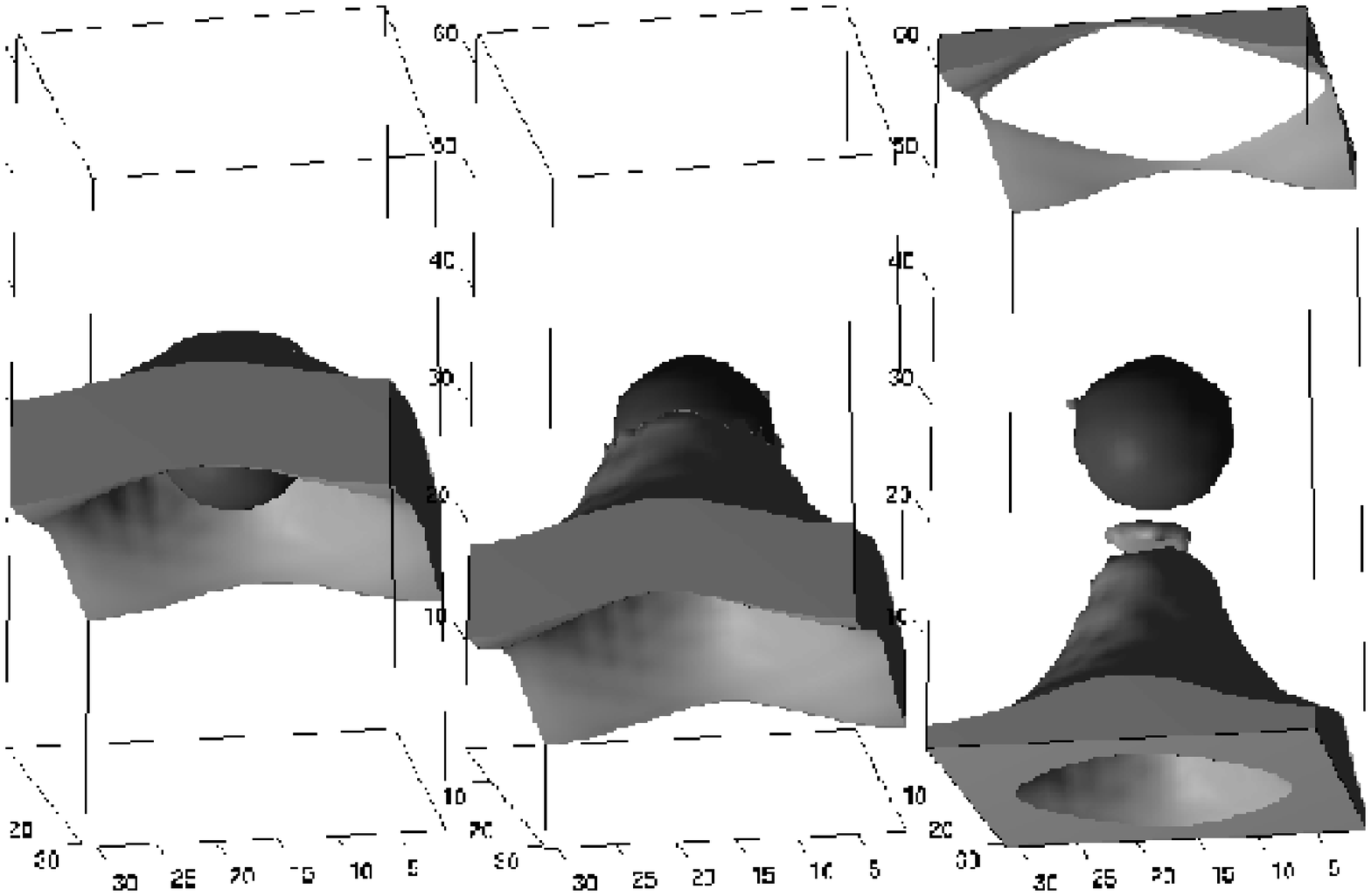}
	
	\caption{
	Convergence of the viscous, 
	low Reynolds number flow of a band of homopolymer B 
	past a circular cylinder in a matrix of homopolymer A under grid refinement.
	Snapshots are taken at $t=0,2,4,6,8,$ and $10$ at three resolutions, 
	(\textbf{a}) 32x64 (red) (\textbf{b}) 48x96 (green) (\textbf{c}) 64x128 (blue). 
	(\textbf{d}) Superposition of homopolymer A interface at all three resolutions at t=6. 
	(\textbf{e}) The same system in three dimensions.}
	\label{fig:flow2d}
\end{figure}

We examine the pressure driven flow past a fixed sphere in greater detail to test the behavior of the hydrodynamic implementation under grid refinement. The velocity field is generated 
by an external force, $\mathbf{f}^\phi=0.2$ which drives a melt of homopolymer $A$ from top to bottom in the channel. The system also contains a transverse band of incompatible homopolymer $B$  used to track the fluid motion. The Flory incompatibility is set at $N\chi_{AB}=8$, and the simulation domain is periodic with dimension $10 \times 20R_g$. The fixed sphere occupies $40\%$ of the channel width and interacts neutrally with both materials. The viscosity is $\eta_s=1.0,$ and viscoelastic effects are suppressed, $G_{0 i}=K_{0 i}=0$.

The interaction of the transverse homopolymer band with the fixed spherical obstacle produces a sequence of events in which the homopolymer band wraps around the sphere and is stretched until it ruptures, leaving a layer of B in contact with the obstacle and a drop of homopolymer B which coalesces in the wake of the obstacle, as shown in fig.~\ref{fig:flow2d}. We repeat the simulation at grids resolutions of 32x64, 48x96, and 64x128,  and compare the volume fraction contours produced by each in fig.~\ref{fig:flow2d}a-d. The contours of homopolymer A are superimposed and are found to be nearly identical, indicating that the method converges to a single solution as the simulation mesh is refined.  We also characterize the flow rate by measuring the maximum Weissenberg number $\mbox{Wi} = \tau v_{\max}/L$ where $v_{\max}$ is the maximum mean field velocity, and L is the width of the narrowest portion of the channel. In all three cases, we obtain the same value, $\mbox{Wi} = 0.32$. In fig.~\ref{fig:flow2d}(e) we analyze this system in three dimensions illustrating that the method produces results consistent with the two dimensional simulations.

\subsection{Viscoelastic Properties}

\begin{figure}[hbt]
	\centering
	{
		\textbf{a} \includegraphics[width=0.95\linewidth]{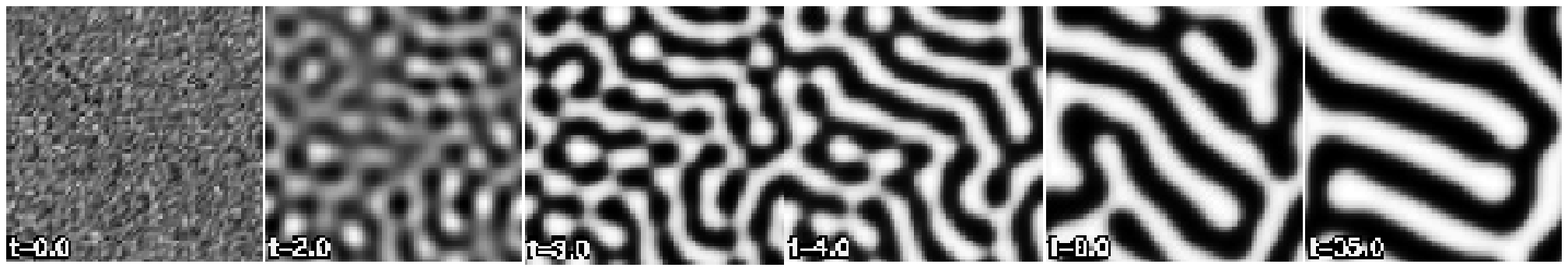}
		
		\textbf{b} \includegraphics[width=0.95\linewidth]{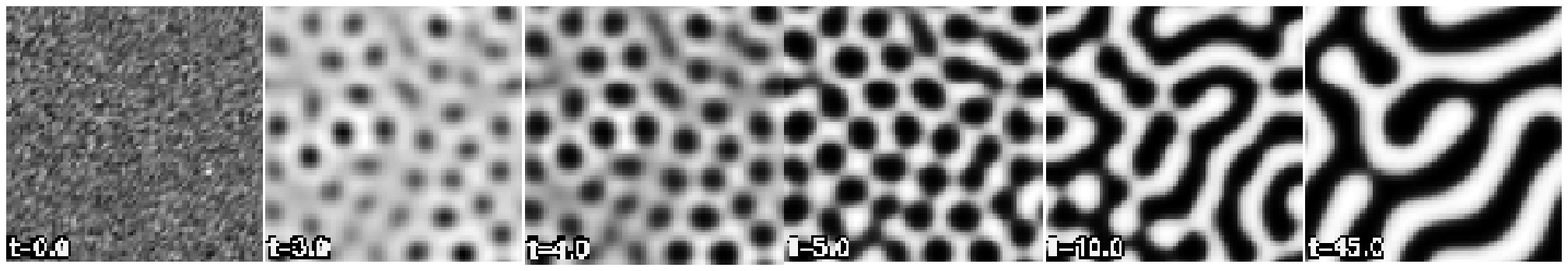}

		\textbf{c} \includegraphics[width=0.95\linewidth]{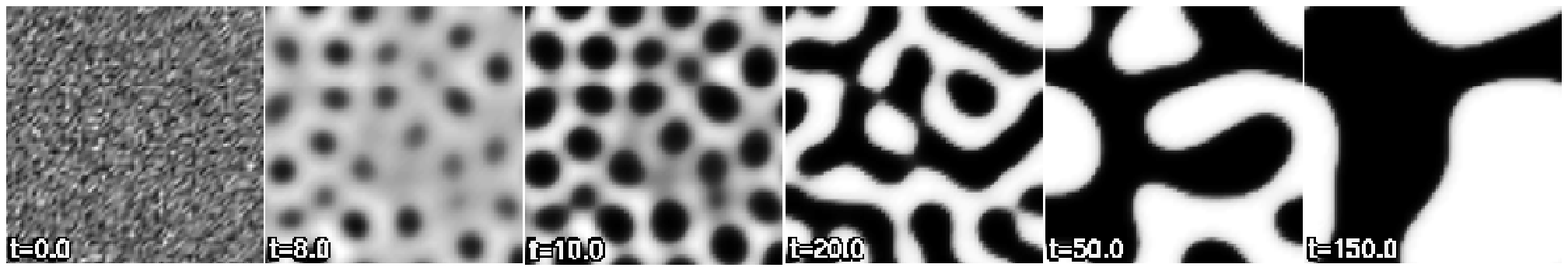}

		\textbf{d} \includegraphics[width=0.95\linewidth]{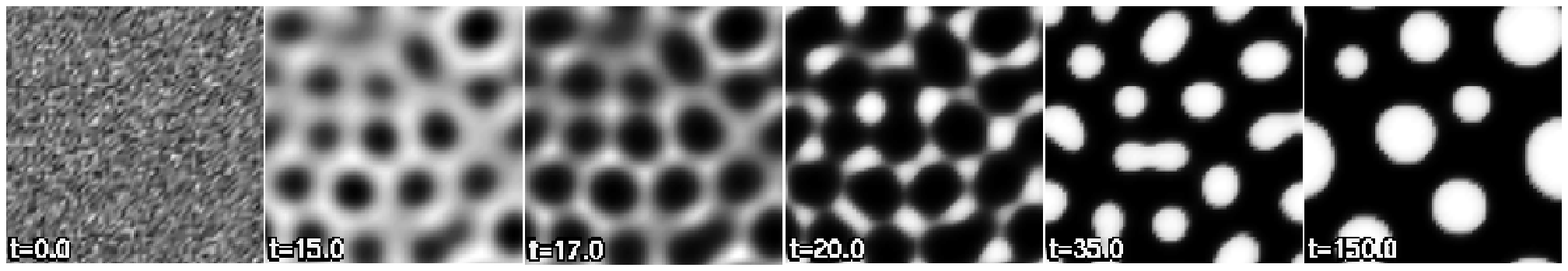}

		\textbf{e} \includegraphics[width=0.95\linewidth]{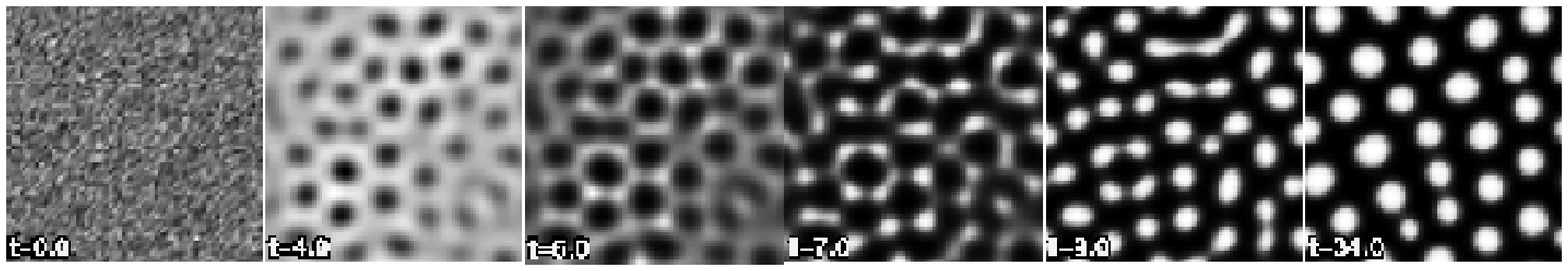}

		\textbf{f} \includegraphics[width=0.95\linewidth]{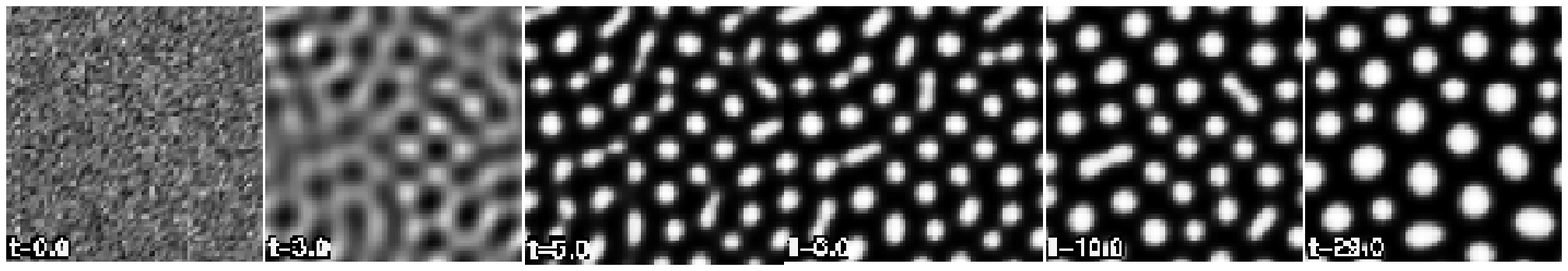}

	}
	\caption{
	Viscoelastic phase separation of two homopolymer components,
	(a) Normal AB diblock, $f_A=0.5$
	(b) Glassy AB diblock, $f_A=0.5$
	(c) Glassy A+B blend, $f_A=0.5$
	(d) Glassy A+B blend, $f_A=0.3$
	(e) Glassy AB diblock, $f_A=0.3$
	(f) Normal AB diblock, $f_A=0.3$
	}
	\label{VISCOELASTIC_RESULTS}
\end{figure}

We validate the viscoelastic components of the model by examining the viscoelastic phase separation of two component systems. ``Viscoelastic phase separation" is the term given to the de-mixing process in two fluids systems which exhibit a large contrast in their dynamic properties (dynamic asymmetry.) In such a system, phase separation occurs in a manner which differs from the usual spinodal decomposition process. Tanaka observed this behavior in polymer solutions \cite{UnusualPhaseSeparation}  and modified the two fluid model to account for them \cite{ViscoelasticPhaseSeparation} . His model has subsequently been applied to viscoelastic phase separation of polymer blends and diblock copolymer melts as well \cite{HuoZhangDiblocks,LuoViscoelasticBlends}.

In fig.~\ref{VISCOELASTIC_RESULTS}, we illustrate the time dependent behavior of six systems which exhibit phase separation behavior. Cases (a) and (f) are included for reference, where case (a) illustrates normal phase separation in a critical diblock copolymer melt, $f_A=0.5$, and case (f) demonstrates phase separation in an off-critical melt $f_A=0.30$. 
Cases (b)-(d) demonstrate viscoelastic phase separation behavior of systems in which a higher bulk modulus material $K_A=5, \zeta_A^\phi=2$ (white) has been mixed with a softer material $K_B=0, \zeta_B^\phi=1$ (black) both of which demonstrate nonzero shear moduli $G_A=G_B=0$. All four systems exhibit the expected characteristic behavior beginning with an incubation period in which the hard phase forms a viscoelastic network structure, followed by nucleation of the softer phase, and eventual network break-up. Case (b) is a critical $f_A=0.5$ glassy/elastic diblock and case (c) is a critical blend. Case (d) is an off-critical ,$f_A=0.3$, glassy/elastic blend and (e) is an off-critical diblock melt. As expected, after the breakup of the network, each material resumes its phase separation processes in which the blends macro-phase separate and the diblocks do not. Cases (d) and (e) demonstrate ``phase inversion" wherein the soft material forms drops in a hard matrix in the early stages and their roles are reversed in the late stages. Each of these results is consistent with previous investigations \cite{HuoZhangDiblocks,LuoViscoelasticBlends}

\section{Conclusions}\label{CONCLUSIONS}

To summarize, we have introduced a method called Hydrodynamic Self-Consistent Field Theory (HSCFT) which extends the capabilities of SCFT to the hydrodynamic regime in which bulk material transport and viscoelastic effects play a significant role. We introduced a semi-implicit, iterative numerical scheme for the solution of these equations which makes liberal use of pseudo-spectral fast-Fourier techniques, enabling practical simulation of sizable systems. We validated the thermodynamic component by reproducing the correct equilibrium meso-phases found in three dimensional diblock copolymer melts.  We validated the hydrodynamic model by demonstrating that it produces low Reynolds number flows that are consistent with analytic solutions of the Stokes flow equations. The viscoelastic constitutive equations were validated by simulating the expected viscoelastic phase separation dynamics of glassy/elastic AB diblock copolymer melts and glassy/elastic A+B blends. We also performed a series of resolution studies on systems with and without bulk flow in which we demonstrated that the combined system is stable and converges smoothly to a single solution under grid-size refinement. We believe that the coupled interaction  of this system of equations provides a powerful and flexible tool for studying the behavior of complex fluids in hydrodynamic flows.

\appendix

\section{Derivation of the Thermodynamic Model}\label{THERMO_DERIVATION}

In this section, we detail the derivation of a non-equilibrium, self consistent field theoretic model 
general enough to represent an arbitrary blend of $C$ distinct multiblock copolymers with varying molecular weights. The derivation follows the standard procedure outlined in \cite{FredricksonSimulationMethods}, which consists of the following steps:
\\
(1) Construct a particulate, mesoscale model of the essential physics.\\
(2) Convert the model to field theoretic form.\\
(3) Approximate the partition function and obtain the thermodynamic forces.

\subsection{Mesoscale Model Construction}

The most general system we are interested in describing consists of $C$ copolymer species and $W$ solid wall materials, which may be visualized as a dense array of mesoscale particles or ``monomers" of equal volume $v_0$. The forces between the monomers consist of a hard core repulsion, covalent bonds between monomers on the same polymer, and Van der Waals forces between non-bonded monomers.

The hard core potential is modeled implicitly by enforcing the incompressibility of the copolymer melt such that the total number density remains constant $ \sum_{i=1}^M \hat{\rho}_i = \rho_0 = 1/v_0$. The sum runs over all distinguishable monomers $M$ and the number density operator for monomer $i$ is
$ \hat{\rho}_{i}(\mathbf{r})= \sum_{j=1}^{n_\alpha} \int_0^{N_\alpha} ds
\left[\delta\left(\mathbf{r} - \mathbf{R}_{\alpha j}(s) \right)\gamma_i(s)\right]$,
where the quantity $n_\alpha$ counts the number of copies of copolymer type $\alpha$ and $N_\alpha$ is its polymerization index.

Covalent bonds are modeled by treating each group of wall particles as a single rigidly rotating and translating object and each copolymer as a string of monomers on a parameterized space curve $\mathbf{R}_{\alpha i}(s)$ where $\alpha\in C$ indicates the copolymer type and $i\in n_\alpha$. Using the standard Gaussian thread approximation, the unperturbed polymer chains are represented as random walks with a ``stretching" free energy given by
\begin{eqnarray}
	U_0 = \sum_{\alpha=1}^{C} \sum_{i=1}^{n_\alpha}
	\frac{kT}{4R_{g\alpha}^2} \int_0^1 ds
	\left| \frac{d\mathbf{R}_{\alpha i}(s) }{ds} \right|^2
\end{eqnarray}
where $R_{g\alpha}$ is the unperturbed radius of gyration of copolymer species $\alpha$.

Van der Waals interactions between non-bonded particles are represented by the interaction potentials
\begin{eqnarray}
	U_{\phi \phi} &=& \frac{kT}{2 v_0}
	\int d\mathbf{r} \sum_{i=1}^M\sum_{j=1}^M
	\hat{\phi}_{i}(\mathbf{r})\chi_{i j} \hat{\phi}_j(\mathbf{r})
	\\
	U_{\phi \psi} &=& \frac{kT}{v_0}
	\int d\mathbf{r} \sum_{i=1}^M\sum_{j=1}^W
	\hat{\phi}_{i}(\mathbf{r})\xi_{i j} \hat{\psi}_j(\mathbf{r})
\end{eqnarray}
where $U_{\phi \phi}$ is the total energy of monomer-monomer interactions, $U_{\phi \psi}$ is the total energy of monomer-wall interactions, and the volume fraction operator  is defined as $\hat{\phi}_i(\mathbf{r}) = \hat{\rho}_i(\mathbf{r})/\rho_o$. Similarly, $\hat{\psi}_j(\mathbf{r})$ represents the volume fraction operator of solid material $j$ at $\mathbf{r}$. The strength of the effective repulsion between dissimilar units is given by the Flory parameters $\chi$ and $\xi$.

We complete the mesoscale model by constructing a partition function over all physically realizable configurations of the system in the canonical ensemble
\begin{eqnarray}
	Z=Z_0 \int \prod_{\alpha,i,j} 
	\mathcal{D}[\mathbf{R}_{\alpha i}]
	\mathcal{D}[\mathbf{r}]\delta(\phi_j(\mathbf{r}) - \hat{\phi}_j(\mathbf{r}))
	\exp(-U/kT)
\end{eqnarray}
where the total energy of a given configuration is $U[\{\mathbf{R}_{\alpha i}\}] = U_0 + U_{\phi \phi} + U_{\phi \psi}$. 
The ensemble sum runs over all possible space curves for each copolymer with delta functions which select the configurations that are consistent with the volume fraction fields $\phi_i(\mathbf{r})$.

\subsection{Conversion to Field Theoretic Form}
The mesoscale model is converted to field theoretic form using a formal procedure which replaces the particle degrees of freedom with a set of continuous volume fraction and chemical potential fields.
Using the delta functions,  the volume fraction operators $\hat{\phi}(\mathbf{r})$ are replaced with their equivalent field values $\phi(\mathbf{r})$ in the interaction terms
\begin{eqnarray}
	U_1/kT &=& \frac{1}{v_0}
	\int d\mathbf{r} \left[
	 \sum_{i=1}^M\sum_{j=1}^M
	\frac{1}{2}\phi_{i}(\mathbf{r})\chi_{i j} \phi_j(\mathbf{r})
	+  \sum_{k=1}^M\sum_{l=1}^W
	\phi_k(\mathbf{r})\xi_{k l} \psi_l(\mathbf{r}) \right]
\end{eqnarray}
and the conjugate chemical fields $\omega_i(\mathbf{r})$ are introduced by employing a Fourier representation for the delta functions. 
\begin{eqnarray}
	\int\mathcal{D}[\mathbf{r}] 
	\delta(\phi_j(\mathbf{r})-\hat{\phi}_j(\mathbf{r})) =
	 \int_{-i\infty}^{i\infty} \mathcal{D}[\omega_j] \exp
	 \left[ \frac{1}{v_0} \int d\mathbf{r} \omega_j(\mathbf{r})[\phi_j(\mathbf{r})
	 - \hat{\phi}_j(\mathbf{r})] \right]
\end{eqnarray}

Because each instance of a copolymer of type $\alpha$ is indistinguishable from others of the same species, the part of $Z$ which depends explicitly on $\mathbf{R}_{\alpha i}$ factors into a product of $n_\alpha$ identical single chain partition functions when we insert the definition of the volume fraction operators $\hat{\phi}$
\begin{eqnarray} 
	Q_\alpha = Q_0 \int \mathcal{D}[\mathbf{R}]
	\exp\left[ 
	-\int_0^1 ds \left(N_\alpha \Omega_\alpha(s)
	+\frac{1}{4R_{g\alpha}^2} \left| \frac{d\mathbf{R}}{ds}\right|^2\right)
	\right]
\end{eqnarray} 
where we have defined $\Omega_\alpha(\mathbf{r},s) = \sum_{i=1}^M\omega_i(\mathbf{r})\gamma_i(s)$.

This integral over all paths may be converted into a volume integral over a propagator in a manner analogous to that employed by Feynman and Kac \cite{FeynmanKac} in their path integral formulation of quantum mechanics. When expressed in this form, $Q_\alpha=\frac{1}{V}\int d\mathbf{r} q_\alpha(\mathbf{r},s)q^\dagger_\alpha(\mathbf{r},s)$, and the propagator and co-propagator  are evaluated by numerically solving the differential equations with initial conditions $q_\alpha(\mathbf{r},0)=1$ and $q^\dagger_\alpha(\mathbf{r},1)=1$.
\begin{eqnarray}
	\partial_s q_\alpha(\mathbf{r},s) &=& 
		+R_{g \alpha}^2 \nabla^2 q_\alpha(\mathbf{r},s)
		- N_\alpha \Omega_\alpha(s) q_\alpha (\mathbf{r},s)\\
	\partial_s q^\dagger_\alpha(\mathbf{r},s) &=& 
		-R_{g \alpha}^2 \nabla^2 q^\dagger_\alpha(\mathbf{r},s)
		+ N_\alpha\Omega_\alpha(s) q^\dagger_\alpha (\mathbf{r},s)
\end{eqnarray}

The partition function $Z$ may now be expressed in the purely field theoretic form
$Z=Z_0 \int \prod_i \mathcal{D}[\omega_i]\exp\left({F[\{\omega,\phi\}]/kT}\right)$ where the quantity F may be interpreted as the free energy associated with a particular configuration of fields $\omega(\mathbf{r})$ and $\phi(\mathbf{r})$ which takes the form
\begin{eqnarray}
	\frac{F}{kT} = \frac{1}{v_0}
	\int \!\!d\mathbf{r}' \left[\sum_{i>j}^M \phi_{i}\chi_{i j} \phi_j
	+  \sum_k^M \sum_l^W\phi_i\xi_{k l} \psi_l - \sum_m^M \phi_m \omega_m\right]
	- \sum_\alpha^C n_\alpha \ln Q_\alpha \nonumber
	\\
\end{eqnarray}

\subsection{Mean Field Approximation}

Since we do not know how to evaluate the partition function, it must be numerically sampled or analytically approximated. As discussed in \cite{FredricksonSimulationMethods}, the chemical potential fields $\omega_i$ are in general complex, which makes direct numerical sampling difficult and time consuming. Instead, a mean field approximation of the partition function is obtained by performing a stationary point analysis on the integral for the conjugate fields $\omega_i(\mathbf{r})$. The resulting solution is purely real and may be obtained from the local thermodynamic equilibrium conditions (LTE)
\begin{eqnarray}
	\frac{\delta F}{\delta \omega_i(\mathbf{r})} =
	 \frac{kT}{v_0}[\phi_i(\mathbf{r}) - \tilde\phi_i(\mathbf{r})]= 0
\end{eqnarray}
The quantitie $\tilde{\phi}_i(\mathbf{r}) = -v_0 n_\alpha \frac{\delta \ln Q_\alpha}{\delta \omega_i(\mathbf{r})}$ is called the auxillary monomer volume fraction, which may be expressed as an integral over the propagators
\begin{eqnarray}
	\tilde{\phi}_i(\mathbf{r})= \frac{ h_\alpha}{Q_\alpha}\int_0^1 ds
 q_\alpha(\mathbf{r},s) q^\dagger_\alpha(\mathbf{r},s) \gamma_i(s)
\end{eqnarray}
where $h_\alpha = n_\alpha N_\alpha (v_0/V)$ is the volume fraction occupied by all copolymers of species $\alpha$ in the system.

The LTE conditions represent a set of highly nonlinear equations
which specify the mean field conjugate potentials $\omega_i(\mathbf{r})$ as a function of the volume fraction fields $\phi_i(\mathbf{r})$. The conjugate fields correspond to local configurations of the copolymer chains which relax quickly in comparison with the conserved volume fraction fields, and as such are be expected to fluctuate close to their mean field values. Once we have solved the LTE conditions, we may compute the non-equilibrium chemical potentials $\mu^\phi(\mathbf{r})$ and $\mu^\psi(\mathbf{r})$ which correspond to functional derivatives of the free energy with respect to the volume fraction fields.
\begin{eqnarray}
	 \mu_i^\phi(\mathbf{r}) &=& \frac{\delta F}{\delta \phi_i(\mathbf{r})} = 
	 \frac{kT}{v_0}\left(
	 \sum_{j=1}^{M}\chi_{ij}\phi_j(\mathbf{r})
	  + \sum_{l=1}^W \xi_{il}\phi_l(\mathbf{r}) -\omega_i (\mathbf{r}) \right)
	  \\
	  \mu_i^\psi(\mathbf{r}) &=& \frac{\delta F}{\delta \psi_i(\mathbf{r})} = 
	 \frac{kT}{v_0}\left(
	 \sum_{j=1}^{M}\xi_{ij}\phi_j(\mathbf{r})\right)
\end{eqnarray}

\section{Derivation of the Multifluid Model}\label{HYDRO_DERIVATION}

In this section, we derive hydrodynamic equations of motion for the transport of multiple viscoelastic fluids in the presence of rigid channel walls using Rayleigh's variational principle as discussed in \cite{DynamicCoupling}. We briefly review the method they employed to derive a two-fluid model for polymer blends and apply the same procedure to derive its multi-fluid generalization.

\subsection{Review of Rayleigh's Variational Principle}

For systems that are not too far out of equilbrium, the theory of irreversible thermodynamics makes the approximation that the generalized velocities in the system are linearly related to the generalized forces by $\frac{dx_i}{dt} = -\sum_j L_{ij} \frac{\partial F}{\partial x_j}$ where $x_i$ are the generalized systems variables, and $L_{ij}$ are Onsager coefficients. For cases where the inverse of the kinetic Onsager coefficient matrix is defined, this relationship may be re-expressed as $\frac{\partial F}{\partial x_i} + \sum_j L^{-1}_{ij}\frac{dx_j}{dt} = 0$. This equation of motion may be equally well expressed in terms of a variational principle in the generalized velocities $\delta R=0$ where $R=\sum_i \frac{\partial F}{\partial x_i}\dot{x}_i + \frac{1}{2}\sum_{i,j}L^{-1}_{ij}\dot{x}_i\dot{x}_j$. The two terms in the Rayleigh functional $R=\dot{F}+\frac{1}{2}W$ may interpreted as the total change in free energy $\dot{F}$ and an energy dissipation function $W$.

\subsection{Multi-fluid Model Derivation}

The total change in free energy for the system under consideration is
\begin{eqnarray}
	\dot{F}=\int d\mathbf{r} \left[ 
	\sum_{i=1}^{M}\frac{\delta F}{\delta \phi_i(\mathbf{r})} \dot{\phi}_i(\mathbf{r}) +
	\sum_{j=1}^W\frac{\delta F}{\delta \psi_j(\mathbf{r})} 
	\dot{\psi}_j(\mathbf{r}) \right]
\end{eqnarray}
In the absence of chemical reactions, the number density of each component is a conserved quantity such that the monomer fields obey $\dot{\phi}_i= - \nabla \cdot \phi_i\mathbf{v}^\phi_i$ and the solid fields obey $\dot{\psi}_j = - \nabla \cdot \psi_j \mathbf{v}^\psi_j$.  Upon substitution of these relationships, the total change in free energy may be written in terms of the monomer velocities $\mathbf{v}^\phi_i$ and the rigid wall velocities $\mathbf{v}^\psi_j$ as
\begin{eqnarray}
	\dot{F}=\int d\mathbf{r} \left[ 
	-\sum_{i=1}^M\mu_i^\phi \nabla \cdot {\phi}_i \mathbf{v}^\phi_i
	-\sum_{j=1}^W \mu_{j}^\psi \nabla \cdot {\psi_j} \mathbf{v}^\psi_j \right] 
\end{eqnarray}

To construct the dissipation function $W$, we note that energy is dissipated due to friction between
the monomers and the entangled polymer network $\zeta_i^\phi(\mathbf{v}^\phi_i - \mathbf{v}_T)^2$, 
by friction between the solid material and the network $\zeta^\psi_j(\mathbf{v}^\psi_j - \mathbf{v}_T)^2$, 
and by elastic deformation of the network $\bm{\sigma}:\nabla \mathbf{v}_T$.
Energy is also added to the system by the motion of the externally driven rigid walls $\mathbf{f}_j^\psi \cdot \mathbf{v}^\psi_j$ and by body forces acting on the bulk of the fluid, $\mathbf{f}_i^\phi \cdot \mathbf{v}_i^\phi$. Therefore, we construct a dissipation function of the form
\begin{eqnarray}
	\!\!\!\!\!\!\!\!\!\!\!
	 \frac{W}{2}\!=\!\!
	 \int \!\!d\mathbf{r}\!
	 \left[
	  \sum_{i=1}^M\!
	  \left(
	  \frac{1}{2}\zeta_i^\phi(\mathbf{v}^\phi_i \!-\! \mathbf{v}_T)^2
	  \!-  \mathbf{f}_i^\phi \cdot \mathbf{v}^\phi_i 
	  \right)
	 +\sum_{j=1}^W  \!\left(
	 \frac{1}{2} \zeta^\psi_j(\mathbf{v}^\psi_j \!-\! \mathbf{v}_T)^2 
	 \!-  \mathbf{f}_j^\psi \cdot \mathbf{v}^\psi_j 
	 \right)
	 +  \bm{\sigma}\!:\!\nabla \mathbf{v}_T\right]
	 \nonumber\\
\end{eqnarray}

The velocity field describing the motion of the entangled polymer network is called the tube velocity 
$\mathbf{v}_T = \sum_i \alpha^\phi_i \mathbf{v}^\phi_i+ \sum_j \alpha^\psi_j \mathbf{v}_j$ which is a rheological mean velocity where each component is weighted by the stress division parameters.
The tube velocity $\mathbf{v}_T$ may also be expressed in terms of the mean velocity $\mathbf{v}$ and the relative velocities $\mathbf{w}_i=\mathbf{v}_i^\phi-\mathbf{v}_T$  using the following relation.
\begin{eqnarray}
	\mathbf{v}_T =\frac{	\sum_i (\alpha_i^\phi-\phi_i)\mathbf{w}_i	
				+ \sum_j (\alpha_j^\psi -\psi_j)\mathbf{v}_i^\psi + \mathbf{v}}
				{1-\sum_k (\alpha_k^\phi-\phi_k)}
\end{eqnarray}

Solving a force balance condition on the entangled network \linebreak[1] 
$\sum_i\zeta^\phi_i(\mathbf{v}^\phi_i-\mathbf{v}_T)
+\sum_j \zeta^\psi_j (\mathbf{v}^\psi_j-\mathbf{v}_T) =0
$
produces the stress division parmeters 
$\alpha^\phi_i=\zeta^\phi_i/\zeta$ and 
$\alpha^\psi_i=\zeta^\psi_j/\zeta$ where the sum of the friction coefficients is
$\zeta = \sum_i^M\zeta^\phi_i + \sum_j^W \zeta^\psi_j$. 
The friction coefficients take the form
$\zeta^\phi_i= \zeta^\phi_{0 i}(N_\alpha/N_{e\alpha})\phi_i$ as discussed in \cite{DynamicCoupling}
 where $N_{e\alpha}$ is the entanglement length of polymer species $\alpha$. The friction coefficient for flow through a semi-porous wall is given by Darcy's law, $\zeta^\psi_j =\zeta^\psi_{0 j}\psi_j/\Phi$ as discussed in \cite{penalizationAgnot,penalizationSchneider}, where the porosity of the material is $\Phi = \sum_i^M\phi_i = 1.0- \sum_j^W \psi_j$.

The equations of motion for each component are obtained by appending the divergence free condition  to the Rayleigh functional $R=\dot{F}+\frac{1}{2}W - p(\nabla \cdot \mathbf{v})$ and then extremizing it with respect to the velocity fields, $\frac{\delta R}{\delta \mathbf{v}^\phi_i}=0$ and $\frac{\delta R}{\delta \mathbf{v}^\psi_j}=0$. The full Rayleigh functional (with implied summation over repeated indices) is
\begin{eqnarray}
	R&=&
	\int \!d\mathbf{r} \left[ 
	-\mu_i^\phi \nabla \cdot {\phi}_i \mathbf{v}^\phi_i
	- \mu_{j}^\psi \nabla \cdot {\psi_j} \mathbf{v}^\psi_j -p(\nabla \cdot \mathbf{v}) 
	+  \bm{\sigma}:\nabla \mathbf{v}_T
	\right. \\
	&&
	\left.
	  +\frac{1}{2} \zeta_i^\phi(\mathbf{v}^\phi_i - \mathbf{v}_T)^2 
	 +\frac{1}{2} \zeta^\psi_j(\mathbf{v}^\psi_j - \mathbf{v}_T)^2 
	 	 -  \mathbf{f}_j^\psi \cdot \mathbf{v}^\psi_j
	 -  \mathbf{f}^\phi_i \cdot \mathbf{v}^\phi_i \right] \nonumber
\end{eqnarray}
which produces the following equations of motion,
\begin{eqnarray}
	0 &=& \phi_i\nabla \mu^\phi_i
	 + \phi_i\nabla p
	  - \alpha^\phi_i \nabla \cdot \bm{\sigma}
	  +\zeta^\phi_i(\mathbf{v}^\phi_i - \mathbf{v}_T) 
	  -\mathbf{f}_i^\phi
	  \label{eqn:eom1}
	  \\
	 0 &=& \psi_j \nabla \mu_{j}^\psi
	 + \psi_j\nabla p
	 - \alpha^\psi_j \nabla \cdot \bm{\sigma}
	 +\zeta^\psi_j(\mathbf{v}^\psi_j - \mathbf{v}_T)
	 -  \mathbf{f}_j^\psi \label{eqn:eom2}
\end{eqnarray}
Eq.~(\ref{eqn:eom1}) may be solved for the velocity of each fluid relative to the elastic polymer network $\mathbf{w}_i = \mathbf{v}^\phi_i - \mathbf{v}_T$, and Eq.(\ref{eqn:eom2}) gives the local forces $\mathbf{f}_j^\psi(\mathbf{r})$ that the walls impart upon the fluid when moving at a velocity $\mathbf{v}^\psi_j(\mathbf{r})$. 
\begin{eqnarray}
	\mathbf{w}^\phi_i  &=&
	\frac{1}{\zeta^\phi_i}\left[
	\alpha^\phi_i \nabla \cdot \bm{\sigma}
	 -\phi_i\nabla \mu^\phi_i
	 - \phi_i\nabla p + \mathbf{f}^\phi_i
	 \right] \\
	\mathbf{f}_j^\psi &=& \psi_j \nabla \mu_{j}^\psi
	 + \psi_j\nabla p
	 - \alpha^\psi_j \nabla \cdot \bm{\sigma}
	 +\zeta^\psi_j(\mathbf{v}^\psi_j - \mathbf{v}_T) \label{eqn:wallForce}
\end{eqnarray}
Summing Eq.~(\ref{eqn:eom1}) and Eq.~(\ref{eqn:eom2}) over all components, 
gives the momentum balance condition 
\begin{eqnarray}
	0 &=& \nabla \pi + \nabla p - \nabla \cdot \bm{\sigma} - \mathbf{f}^\psi -\mathbf{f}^\phi
\end{eqnarray}
in the limit of zero Reynolds number, which together with the divergence free condition implicitly specifies $\mathbf{v}$. The total osmotic pressure is defined as $\nabla \pi = \sum_i^M \phi_i\nabla \mu^\phi_i + \sum_j^W \psi_j \nabla \mu^\psi_j$. The net external force exerted by the walls is $\mathbf{f}^\psi = \sum_j^W \mathbf{f}_j^\psi$ and the total body force acting on the fluid is $\mathbf{f}^\phi =\sum_i^M\mathbf{f}_i^\phi$.

\end{document}